\documentclass[12pt]{spieman}  
\usepackage{amsmath,amsfonts,amssymb}
\usepackage{graphicx}
\usepackage{setspace}
\usepackage{tocloft}
\usepackage{mathtools}

\usepackage{algorithm}
\usepackage{algorithmicx}
\usepackage{algpseudocode}

\usepackage{lineno}

\title{Spatiotemporal Image Reconstruction to Enable High-Frame Rate Dynamic Photoacoustic Tomography with Rotating-Gantry Volumetric Imagers}

\author[a, \textdagger]{Refik Mert Cam}
\author[b, \textdagger]{Chao Wang}
\author[c]{Weylan Thompson}
\author[c]{Sergey A. Ermilov}
\author[d]{Mark A. Anastasio}
\author[e, *]{Umberto Villa}
\affil[a]{University of Illinois Urbana-Champaign, Department of Electrical \& Computer Engineering, 306 North Wright Street, Urbana, IL, United States, 61801}
\affil[b]{Department of Statistics and Data Science, National University of Singapore, Singapore}
\affil[c]{PhotoSound Technologies Inc., Houston, TX 77036, USA}
\affil[d]{University of Illinois Urbana-Champaign, Department of Bioengineering, 1406 West Green Street, Urbana, IL, United States, 61801}
\affil[e]{The University of Texas at Austin, Oden Institute for Computational Engineering \& Sciences, 201 East 24th Street, Austin, TX, United States, 78712}

\cftpagenumbersoff{figure}
\cftpagenumbersoff{table} 
\begin{document} 
\maketitle

\begin{abstract}

\textbf{Significance}: Dynamic photoacoustic computed tomography (PACT) is a valuable imaging technique for monitoring physiological processes. However, current dynamic PACT imaging techniques are often limited to two-dimensional (2D) spatial imaging. While volumetric PACT imagers are commercially available, these systems typically employ a rotating measurement gantry in which the tomographic data are sequentially acquired, as opposed to being acquired simultaneously at all views.  Because the dynamic object varies during the data-acquisition process, the sequential data-acquisition process poses substantial challenges to image reconstruction associated with data incompleteness. 
The proposed image reconstruction method is highly significant in that it will address these challenges and enable volumetric dynamic PACT imaging with existing preclinical imagers.



\textbf{Aim}: The aim of this study is to develop a spatiotemporal image reconstruction method for dynamic PACT that can be applied to commercially available volumetric PACT imagers that employ a sequential scanning strategy. The proposed reconstruction method aims to overcome the challenges
caused by the limited number of tomographic measurements acquired per frame.

\textbf{Approach}: A low-rank matrix estimation-based spatiotemporal image reconstruction method (LRME-STIR) is proposed to enable dynamic volumetric PACT. The LRME-STIR method leverages the spatiotemporal redundancies in the dynamic object to accurately reconstruct a four-dimensional (4D) spatiotemporal image.

\textbf{Results}: The conducted numerical studies substantiate the LRME-STIR method's efficacy in reconstructing 4D dynamic images from tomographic measurements acquired with a rotating measurement gantry. 
The experimental study demonstrates the method's ability to faithfully recover the flow of a contrast agent at a frame rate of 0.1 seconds, even when only a single tomographic measurement per frame is available.



\textbf{Conclusions}: The proposed LRME-STIR method offers a promising solution to the challenges faced by enabling 4D dynamic imaging using commercially available volumetric PACT imagers. By enabling accurate spatiotemporal image reconstructions, this method has the potential to significantly advance preclinical research and facilitate the monitoring of critical physiological biomarkers.

\end{abstract}

\keywords{Photoacoustic computed tomography, Optoacoustic tomography, Spatiotemporal image reconstruction, Dynamic imaging, Small animal imaging, Low-rank matrix estimation}

{\noindent \footnotesize\textbf{\textdagger}Refik Mert Cam and Chao Wang are equal contributors to this work and designated as co-first authors.}

{\noindent \footnotesize\textbf{*}Umberto Villa, \linkable{uvilla@oden.utexas.edu} }


\section{Introduction}
\label{sect:intro}  








Photoacoustic computed tomography (PACT), also referred to as optoacoustic tomography, is an emerging and promising imaging modality with broad applications in the field of biomedical imaging \cite{wang2003noninvasive, brecht2009whole, yang2007functional}. By combining the high spatial resolution of ultrasound imaging with the high soft tissue contrast of optical imaging, PACT offers unique advantages for imaging biological structures while avoiding the use of ionizing radiation. 
In PACT, a fast laser pulse in the near infrared range illuminates the object. Absorption of optical energy by various molecules within the object (chromophore) induces a localized increase of acoustic pressure through the photoacoustic effect. The acoustic wavefield propagating through the object and coupling medium (water) is subsequently detected by ultrasonic  transducers. 
The measured wavefield data can then be utilized to reconstruct an image that depicts the initial induced pressure distribution within the object.

In preclinical and clinical research, the ability to monitor dynamic physiological processes is of utmost importance for comprehending the progression of diseases and developing new treatments \cite{kagadis2010vivo, loudos2011current, franc2008small}. For example, tumor vascular perfusion is one such dynamic processes that is critical in the study of cancers. High vascular perfusion is indicative of angiogenesis, a well-established hallmark of cancerous growth\cite{carmeliet2005angiogenesis, carmeliet2000angiogenesis}. Due to its noninvasive nature and the combination of optical contrast and spatial resolution at depths beyond the optical diffusion limit, PACT represents a promising imaging modality for monitoring critical dynamic physiological processes in preclinical and clinical research \cite{small_animal_pact, li2017single,upputuri2017dynamic, shi2019thermosensitive, manohar2019current}.
Despite its considerable promise, current dynamic PACT technologies suffer from fundamental limitations. They often target two-dimensional (2D) spatial imaging due to shorter data acquisition times and computationally less demanding image reconstruction compared to 3D imaging \cite{lin2018single, gamelin2009real, he2017plasmonic, upputuri2017high, upputuri2017dynamic, shan2020vivo}. Most existing 3D PACT imagers developed to-date  utilize a rotating measurement geometry
in which the tomographic data are sequentially acquired \cite{brecht20183d, brecht2009whole, lin2021high, ermilov2009laser}, as opposed to
being acquired simultaneously at all views. This  design is advantageous because it reduces system costs by employing a limited number of acoustic transducers and associated electronics.  However, data-acquisition times for a complete tomographic scan can be tens of seconds.
  Due to the relatively slow rotational speed, the temporal resolution is significantly limited. While enhancing temporal resolution by using sparsely sampled tomographic data is possible, the associated dynamic image reconstruction problem becomes ill-posed and highly challenging.






Previous studies on dynamic PACT\cite{lin2018single, gamelin2009real, he2017plasmonic, upputuri2017high, upputuri2017dynamic, shan2020vivo, li2017single, 4d_pact}, have primarily focused on scenarios where the sufficiently sampled tomographic data can be rapidly acquired. In such cases, a straightforward approach is to employ a frame-by-frame image reconstruction (FBFIR) method\cite{lin2018single, gamelin2009real, he2017plasmonic, upputuri2017high, upputuri2017dynamic, shan2020vivo, li2017single}. These techniques utilize conventional static image reconstruction methods to estimate a sequence of images from sufficiently sampled tomographic data. The temporal resolution is limited by the duration of the complete data acquisition process. Rapid data acquisition is feasible either with 2D PACT imaging or by leveraging a dense, albeit expensive, static transducer array in 3D imaging. However, for volumetric imagers with sequential scanning strategy, FBFIR methods are not applicable, primarily due to the extended time required to accumulate the complete set of tomographic measurements.


On the other hand, spatiotemporal image reconstruction (STIR) methods estimate a sequence of images simultaneously instead of frame-by-frame, and they have demonstrated their efficacy in accurately reconstructing dynamic objects from sparsely sampled data in various medical imaging modalities, including computed tomography\cite{cai2014cine}, positron emission tomography\cite{spatiotemporal_pet}, single photon emission computed tomography\cite{ding2017dynamic}, and magnetic resonance imaging\cite{haldar2010spatiotemporal}. While a few STIR techniques have been proposed for PACT, some of them still presume the availability of sufficiently sampled tomographic data and strive for enhanced accuracy and/or reduced computational complexity compared to FBFIR techniques\cite{wang2014fast}. The STIR methods\cite{arridge2016accelerated, compressed_razansky, lucka2018enhancing} considering sparsely sampled tomographic data, have relied on the principles of compressed sensing and required specific data sampling schemes that are different than the sequential sampling schemes employed by currently available volumetric imagers.




This study introduces a novel spatiotemporal image reconstruction method based on low-rank matrix estimation (LRME-STIR), which is applicable to currently available sequential 3D PACT imaging systems without requiring hardware modifications. By employing the LRME-STIR technique, it becomes possible to overcome the challenges caused by the sparsely sampled tomographic data. The proposed approach holds the potential to advance the field by enabling accurate and efficient spatiotemporal image reconstruction, thereby facilitating the monitoring of dynamic physiological changes using PACT.

The remainder of the paper is organized as follows. Section \ref{imaging_model} presents an imaging model for sequential volumetric imagers and introduces the inverse problem formulation. The proposed LRME-STIR method is described in Section \ref{lrme_stir}. The conducted numerical and experimental studies, and the results are provided in Sections \ref{study_description} and \ref{results}, respectively. Finally, the paper concludes with a discussion in Section \ref{conclusion}.

\section{Imaging Model for Sequential Volumetric Imagers and Inverse Problem Formulation}\label{imaging_model}







\begin{figure}[tbh]
\centering
\includegraphics[width=0.95\linewidth]{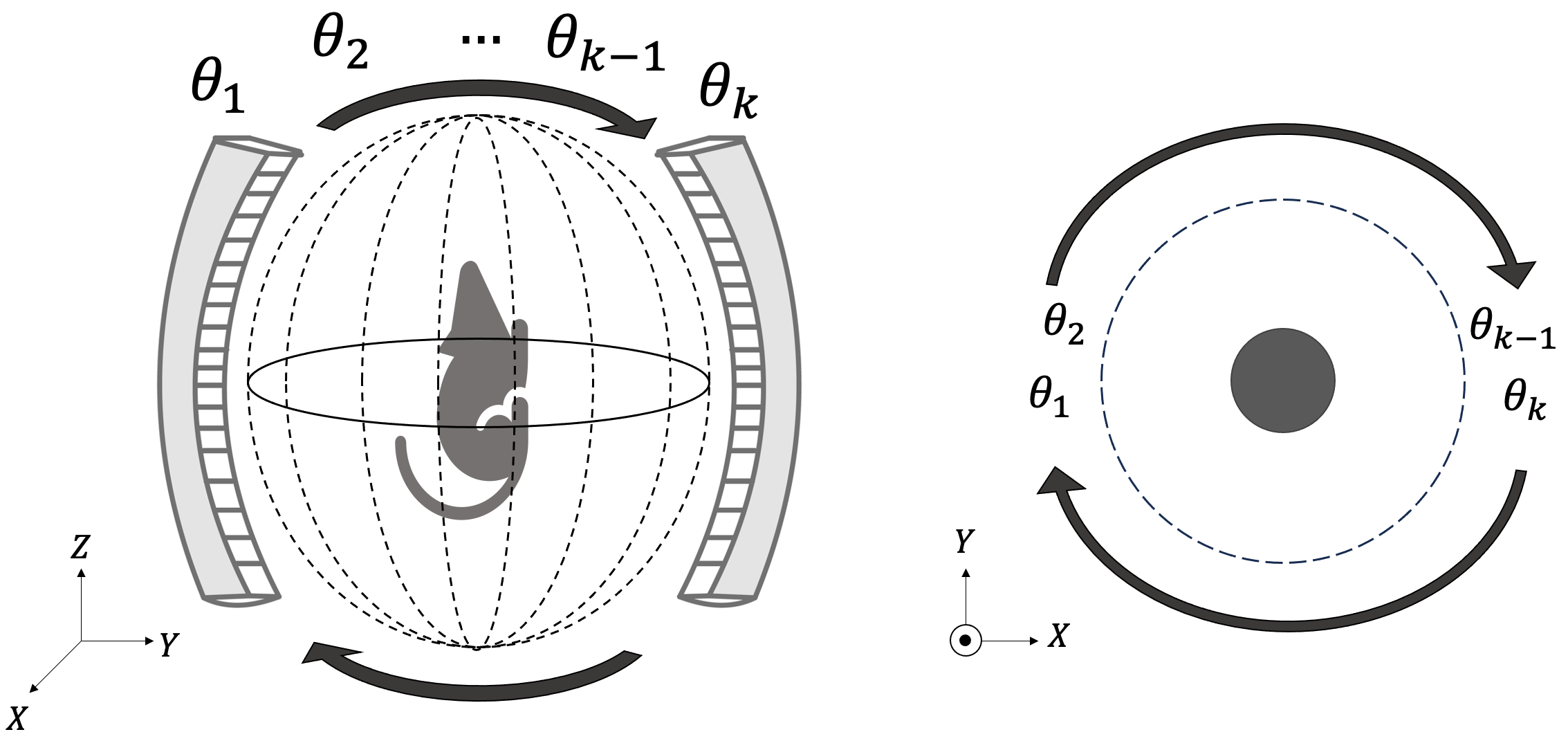}
\caption{Schematic illustrating the sequential scanning strategy employing a rotating acoustic probe around the object.}
\label{sequential_scanning}
\end{figure}

In the context of PACT, the sequential data acquisition strategy commonly involves utilizing one or more rotating or translating ultrasonic transducer arrays within single or multiple acoustic probes\cite{brecht20183d, brecht2009whole, lin2021high}. This approach facilitates data collection by rotating the probes along a fixed axis, resulting in the acquisition of a few tomographic measurements at each step, as depicted in Figure \ref{sequential_scanning}. These measurements are accumulated sequentially to form a complete tomographic measurement set. 

 During each step of sequential data acquisition, the object can be considered as static (quasi-static assumption), which is justified by the negligible data acquisition time during each step (on the order of $10^{-4}$ s), significantly shorter than the time between consecutive measurements (on the order of $10^{-1}$ s). An object frame is defined as the short period of time when the object is considered static. The sequence of object frames constitutes the dynamic object. Essentially, each data acquisition step, hereafter referred to as an imaging frame, corresponds to an object frame. Although object and imaging frames may seem interchangeable, the distinction lies in the fact that the set of imaging frames is essentially a subset of the set of object frames because the dynamic object might not be imaged throughout all object frames. Under the quasi-static assumption, the time-dependent object function at the $k$-th imaging frame, specifically the dynamic induced initial pressure distribution, can be represented as $f_k(\pmb{r}) = f(\pmb{r}, k\Delta t)$ for $k=1,...,K$. Here, $K$ represents the number of imaging frames, and $\Delta t$ is the time interval between consecutive imaging frames, which is equal to the laser repetition rate. Finally, the rotation speed of the gantry determines the angular spacing between imaging frames.

Under the quasi-static assumption, the data acquisition process at each imaging frame can be described by a continuous-to-discrete (C-D) imaging model as\cite{wang2013accelerating, wang2014fast}:
\begin{equation}
[\boldsymbol{g}_k]_{(q-1)P+p} = \frac{1}{\Omega_{qk}}\int_{\Omega_{qk}}d\boldsymbol{r}_{qk}^{\prime}\frac{1}{4 \pi} \int_V d \boldsymbol{r}\frac{f_k(\boldsymbol{r})}{c_0^2}\frac{d}{d\tau}\frac{\delta(\tau-\frac{|\boldsymbol{r}_{qk}^{\prime}-\boldsymbol{r}|}{c_0})}{|\boldsymbol{r}_{qk}^{\prime}-\boldsymbol{r}|}\Bigg|_{\tau=p\Delta T}, \ \genfrac{}{}{0pt}{}{p=1, 2, \dots, P}{q=1, 2, \dots, Q,}
\label{eq:cd-imaging-operator}
\end{equation}
where $\tau$ denotes the fast-time (i.e., the arrival time of acoustic signals to the transducer), and $\Delta T$ corresponds to the fast-time  sampling interval. The object function at the $k$-th imaging frame, $f_k(\boldsymbol{r})$, is assumed to be bounded and contained within the volume $V$. The scalar $c_0$ denotes the speed of sound, which is assumed to be constant throughout the volume $V$. The quantities $\boldsymbol{r}$ and $\boldsymbol{r}_{qk}^{\prime}$ specify the spatial coordinates within $V$ and the location of the $q$-th transducer at the $k$-th imaging frame, respectively. The vector $\boldsymbol{g}_k \in \mathbb{R}^{PQ}$ represents lexicographically ordered pressure traces measured by the transducers  at the $k$-th imaging frame. Here, $Q$ stands for the number of transducers, and $P$ represents the number of electrical signals recorded by each transducer. The notation $[\boldsymbol{g}_k]_{(q-1)P+p}$ refers to the $(q-1)P+p$-th entry of the measurement vector $\boldsymbol{g}_k$. Here, the integer-valued indices $q$ and $p$ denote the transducer index and temporal sample, and $\Omega_{qk}$ denotes the detection area of the $q$-th transducer at the $k$-th imaging frame. When the transducer size is small and/or the object is located near the center of a relatively large measurement geometry, an idealized point-like transducer model can be assumed and the surface integral over $\Omega_{qk}$ can be neglected\cite{wang2014fast}.


To facilitate the implementation of an iterative image reconstruction algorithm, a discrete-to-discrete (D-D) imaging model is defined as follow. The spatially continuous object functions $f_k$ corresponding to the $k$-th imaging frame are  approximated by use of a finite linear combination of  spatial expansion functions $\{\psi_n(\boldsymbol{r})\}_{n=1}^N$:
\begin{equation}
f_k^{(N)}(\boldsymbol{r}) = \sum_{n=1}^N \alpha_n^k \psi_n(\boldsymbol{r}), \quad k=1,\dots,K,
\label{eq:object-function-expansion}
\end{equation}
where $N$ denotes the number of spatial expansion functions.
In this study, the expansion functions are piecewise trilinear Lagrangian functions defined on a uniform Cartesian grid. Their expression is given by\cite{wang2013accelerating}:
\begin{equation}
  \psi_n(\boldsymbol{r}) =
    \begin{cases}
      \left(1-\frac{|x-x_n|}{\Delta s}\right) \left(1-\frac{|y-y_n|}{\Delta s}\right) \left(1-\frac{|z-z_n|}{\Delta s}\right),& \text{if $|x-x_n|$, $|y-y_n|$, $|z-z_n| \leq \Delta s$}\\
      0, & \text{otherwise.}
    \end{cases}       
\end{equation}
Here, $\boldsymbol{r} = (x,y,z)$ denotes the spatial coordinate and $\boldsymbol{r}_n = (x_n, y_n, z_n)$ specifies the location of the $n$-th node of the uniform Cartesian grid. The parameter $\Delta s$ indicates the distance between adjacent grid points.

The coefficients $\{\alpha_n^k\}_{n=1,k=1}^{NK}$ can be organized into a matrix $\boldsymbol{F} \in \mathbb{R}^{N\times K}$ with entries $[\boldsymbol{F}]_{nk} \equiv \alpha_n^k$. The $k$-th column of $\boldsymbol{F}$ is denoted by $\boldsymbol{f}_k \in \mathbb{R}^N$ and represents the discrete approximation of the object function at the $k$-th imaging frame. That is,
\begin{equation}
\boldsymbol{F} = 
\sum_{k=1}^K \boldsymbol{f}_k \otimes \boldsymbol{e}_k =
\begin{bmatrix}
    \alpha_1^1 & \dots & \alpha_1^K \\
    \vdots & \ddots & \vdots\\
    \alpha_N^1 & \dots & \alpha_N^K 
    \end{bmatrix},
\end{equation}
where $\boldsymbol{e}_k \in \mathbb{R}^K$ represents the $k$-th column of the identity matrix in $\mathbb{R}^K$ and $\otimes$ denotes the vector outer product.
Correspondingly, a frame-dependent D-D imaging model accounting for  measurement noise can be expressed as:
\begin{equation}
\underline{\boldsymbol{g}}_k = \boldsymbol{H}_k \boldsymbol{f}_k + \boldsymbol{\eta}_k, \quad k=1, 2, \dots K, 
\end{equation}
where $\underline{\boldsymbol{g}}_k$ represents the (noisy) tomographic measurements at the $k$-th imaging frame and $\boldsymbol{\eta}_k$ accounts for measurement noise and modeling and discretization errors. The operator $\boldsymbol{H}_k \in \mathbb{R}^{QP\times N}$ stems from discretization of the C-D PACT imaging operator associated with the $k$-th imaging frame. In particular, the vector $\boldsymbol{g}_k \in \mathbb{R}^{QP}$ representing the action of $\boldsymbol{H}_k$ on $\boldsymbol{f}_k$ is defined as in Eqn. \eqref{eq:cd-imaging-operator} but with  the continuous in space object function $f_k(\boldsymbol{r})$ replaced  by its finite dimensional approximation $f_k^{(N)}(\boldsymbol{r})$ defined in Eqn. \eqref{eq:object-function-expansion}.

Given the data matrix $\underline{\boldsymbol{G}}=\sum_{k=1}^K \underline{\boldsymbol{g}}_k \otimes \boldsymbol{e}_k \in \mathbb{R}^{PQ \times K}$ and the set of imaging operators $\boldsymbol{H_k}$ ($k=1, \ldots, K$) corresponding to each imaging frame, the goal of dynamic PACT image reconstruction is to find a matrix $\boldsymbol{\hat{F}}=\sum_{k=1}^K
\boldsymbol{\hat{f}}_k \otimes \boldsymbol{e}_k \in \mathbb{R}^{N\times K}$, whose column $\boldsymbol{\hat{f}}_k$ represents an object estimate at the $k$-th imaging frame; for simplicity hereafter $\boldsymbol{\hat{f}}_k$ is referred to as the $k$-th frame of the spatiotemporal object estimate. Given the sparse tomographic measurements acquired for each imaging frame, the task of dynamic image reconstruction constitutes a significantly ill-posed inverse problem. A unique estimator, $\boldsymbol{\hat{F}}$, can be obtained by solving the following penalized least squares optimization problem:
\begin{equation}
    \boldsymbol{\hat{F}} = \operatornamewithlimits{argmin}\limits_{\boldsymbol{F} \in \mathbb{R}^{N \times K}} \text{\textbf{J}}(\boldsymbol{F}) \coloneqq \text{\textbf{L}}(\boldsymbol{F})+\text{\textbf{R}}(\boldsymbol{F}) = \sum_{k=1}^K \text{\textbf{L}}_k(\boldsymbol{F})+\text{\textbf{R}}(\boldsymbol{F}),
\end{equation}
where $\text{\textbf{R}}(\boldsymbol{F})$ is the regularization term, which is convex but possibly non-smooth. The total data fidelity term $\text{\textbf{L}}(\boldsymbol{F}) = \sum_{k=1}^K \text{\textbf{L}}_k(\boldsymbol{F})$ is the sum of  data fidelity terms  $\text{\textbf{L}}_k(\boldsymbol{F})$ associated with the $k$-th imaging frame. These quantities are defined as
\begin{equation}
\text{\textbf{L}}(\boldsymbol{F}) =  \frac{1}{2}\left\lVert \left(\sum_{k=1}^K (\boldsymbol{H}_{k}\boldsymbol{f}_{k}) \otimes \boldsymbol{e}_k  \right) - \underline{\boldsymbol{G} }\right\rVert^2_F \quad \text{and} \quad \text{\textbf{L}}_k(\boldsymbol{F}) = \frac{1}{2}\left\lVert\boldsymbol{H}_{k}\boldsymbol{f}_{k}-\boldsymbol{g}_{k}\right\rVert^2_2,
\label{eq:framewise-data-fidelity}
\end{equation}
respectively, where $\lVert \cdot \lVert_F$ denotes the Frobeniuos norm.

In addition to the inherent ill-posed nature of the considered dynamic image reconstruction problem, significant implementation challenges exist. Firstly, unlike the relatively straightforward task of static image reconstruction in which a single image is estimated, spatiotemporal image reconstruction involves dealing with a considerably higher computational burden as all frames are reconstructed simultaneously, particularly when each frame is 3D in space. Secondly, as the number of frames increases, there is a growing need for memory space during computation, which necessitates effective computational strategies with minimal memory usage.

\section{Low-rank Matrix Estimation-based Spatiotemporal Image Reconstruction}\label{lrme_stir}
In numerous biomedical applications, the spatiotemporal object function of interest has been demonstrated to be effectively approximated by a small set of weights $\sigma_r$, and functions $u_r(\pmb{r})$ and $v_r(t)$, depending only on the space or time variables, known as the semiseparable approximation\cite{liang2007spatiotemporal, brinegar2009real, cam2023dynamic, cam2023low, lozenski2022memory, haldar2010spatiotemporal, lingala2011accelerated}. Consequently, the object function can be expressed as $f(\pmb{r},t) \approx \sum_{r=1}^{R} \sigma_r u_r(\pmb{r})v_r(t)$ \cite{liang2007spatiotemporal, brinegar2009real, cam2023dynamic, cam2023low, lozenski2022memory, haldar2010spatiotemporal, lingala2011accelerated}. Leveraging the semiseparable approximation, the spatiotemporal reconstruction problem can be reduced to the problem of estimating $R$ weights, and $R$ spatial and temporal functions. In a discretized formulation, this is algebraically equivalent to enforcing a low-rank structure on the matrix $\boldsymbol{F}$.

A penalty scheme can then be imposed on the nuclear norm of the spatiotemporal reconstruction matrix to promote low-rankness. Specifically, the regularization term stemming from the nuclear norm of $\boldsymbol{F}$ is defined as
\begin{equation}
\text{\textbf{R$^{nn}$}}(\boldsymbol{F}) =  \| \boldsymbol{F} \|_* = \sum_{r=1}^{\min(N,K)} \sigma_r,
\label{eq:nuclear-norm}
\end{equation}
where $\sigma_r$ ($r=1, \ldots, \min(N,K)$) represent the singular values of $\boldsymbol{F}$.
This approach not only effectively regularizes the ill-posed inverse problem \cite{gu2014weighted} but also reduces memory demands without sacrificing accuracy. Rather than explicitly storing $\boldsymbol{F}$  in memory, its truncated singular value decomposition $\boldsymbol{U}_R\boldsymbol{\Sigma}_R\boldsymbol{V}_R^T$ is stored, resulting in decreased memory requirement. Here, $R < \min(N,K)$ denotes the truncation index,
$\boldsymbol{\Sigma}_R$ is the diagonal matrix comprising the largest $R$ singular values,
$\boldsymbol{U}_R$ and $\boldsymbol{V}_R$ are matrices with orthonormal columns collecting the  left and right singular vectors, respectively, corresponding to the $R$ largest singular vaslues.
This approach not only facilitates an efficient approximation of $\boldsymbol{F}$ but also enforces the space-time semiseparability. Specifically, the columns of $\boldsymbol{U}_R$ and $\boldsymbol{V}_R$ are the algebraic counterpart of the functions $u_r(\pmb{r})$ and $v_r(t)$, respectively.




To further regularize the inverse problem, under the assumption that the object undergoes a smooth and slow temporal change, another regularization scheme penalizing the difference between two consecutive frames can be employed. This can be accomplished by penalizing the squared Frobenius norm of the temporal difference matrix, which is expressed through the temporal (forward) difference operator, denoted as $\boldsymbol{D} \in \mathbb{R}^{K \times (K-1)}$, applied to $\boldsymbol{F}$:
\begin{equation}
\text{\textbf{R$^{t}$}}(\boldsymbol{F}) = \frac{1}{2} \left\lVert \boldsymbol{F} \boldsymbol{D} \right\rVert^2_F,
\label{eq:temporal-tikhonov}
\end{equation}
where, $\boldsymbol{D}$ is defined as:
\begin{equation}\label{temporal_diff_op}
\boldsymbol{D} = \big[\begin{array}{c|c|c|c|c}
   \boldsymbol{d}_{1} &\cdots &\boldsymbol{d}_{k} &\cdots &\boldsymbol{d}_{K-1}
   \end{array}\big] = 
\begin{bmatrix}
    -1 & 0 & 0 & 0 & \dots \\
    1 & -1 & 0 & 0 & \dots \\
    0 & 1 & -1 & 0 & \dots \\
    \vdots & \vdots & \vdots & \vdots & \vdots  \\
    0 & \dots & 0 & 0 & -1 \\
    0 & \dots & 0 & 0 & 1
    \end{bmatrix}.
\end{equation}
Here, $\boldsymbol{d_k}$ corresponds to the $k$-th column of the temporal (forward) difference operator, $\boldsymbol{D}$. Within this column, the $k$-th and $(k+1)$-th elements possess values of $-1$ and $1$, respectively, while all other elements are set to $0$. 

In this way, the sought after estimate of the dynamic reconstruction problem with consideration of a maximum rank constraint and temporal and nuclear norm penalties can be formulated as: 
\begin{equation}\label{lrme-stir-opt}
\begin{aligned}
\hat{\boldsymbol{F}} &= \operatornamewithlimits{argmin}\limits_{\boldsymbol{F} \in \mathbb{R}^{N \times K}_{R_{\max}} }\text{\textbf{J}}(\boldsymbol{F}) \coloneqq\text{\textbf{L}}(\boldsymbol{F})+ \gamma\text{\textbf{R$^{t}$}}(\boldsymbol{F})+\lambda\text{\textbf{R$^{nn}$}}(\boldsymbol{F}),
\end{aligned}
\end{equation}
where $\mathbb{R}^{N \times K}_{R_{\max}}$ denotes the set of all $N$-by-$K$ matrices of rank at most $R_{\max}$ and the cost function $\text{\textbf{J}}(\boldsymbol{F})$ is comprised of the data fidelity term $\text{\textbf{L}}(\boldsymbol{F})$ in Eqn. \eqref{eq:framewise-data-fidelity}, the temporal regularization term $\text{\textbf{R$^{t}$}}(\boldsymbol{F})$ in Eqn. \eqref{eq:temporal-tikhonov}, and  the nuclear norm regularization term $\text{\textbf{R$^{nn}$}}(\boldsymbol{F})$ in Eqn. \eqref{eq:nuclear-norm}. Here, the parameters $\gamma \geq 0$ and  $\lambda \geq 0$ control the strength of the temporal and nuclear norm regularization terms, respectively. 
Similarly to the data fidelity computation in Eqn. \eqref{eq:framewise-data-fidelity}, the temporal regularization term $\text{\textbf{R$^{t}$}}(\boldsymbol{F}) = \sum_{k=1}^K \text{\textbf{R$_k^{t}$}}(\boldsymbol{F})$ can written as the sum of contributions $\text{\textbf{R$_k^{t}$}}$ from each imaging frame. Each term is defined as
\begin{equation}
\text{\textbf{R$_k^{t}$}}(\boldsymbol{F}) \coloneqq \begin{cases}
\left\lVert \boldsymbol{F} \boldsymbol{d}_k \right\rVert^2_2 = \lVert\boldsymbol{f}_{k+1}-\boldsymbol{f}_{k}\rVert^2_2 & \text{if } k < K\\
\left\lVert \boldsymbol{F} \boldsymbol{d}_K \right\rVert^2_2 = 0 & \text{if } k = K,\\
\end{cases}
\end{equation}
where, for uniformity of notation, $\boldsymbol{d}_K \in \mathbb{R}^K$ is the zero vector.
To highlight the contribution from each frame, the minimization problem in Eqn. \eqref{lrme-stir-opt} can then be rewritten as
\begin{equation}\label{lrme-stir-opt-by-frame}
\begin{aligned}
\hat{\boldsymbol{F}} &= \operatornamewithlimits{argmin}\limits_{\boldsymbol{F} \in \mathbb{R}^{N \times K}_{R_{\max}} }\sum_{k=1}^{K}\text{\textbf{L}$_k$}(\boldsymbol{F}) + \gamma\sum_{k=1}^{K} \text{\textbf{R$^{t}_k$}}(\boldsymbol{F}) + \lambda \text{\textbf{R$^{nn}$}}(\boldsymbol{F})\\
&= \operatornamewithlimits{argmin}\limits_{\boldsymbol{F} \in \mathbb{R}^{N \times K}_{R_{\max}} }\sum_{k=1}^{K}\lVert\boldsymbol{H}_{k}\boldsymbol{f}_{k}-\boldsymbol{g}_{k}\rVert^2_2 + \frac{\gamma}{2}\sum_{k=1}^{K} \left\lVert \boldsymbol{F} \boldsymbol{d}_k \right\rVert^2_2 +\lambda\lVert\boldsymbol{F}\rVert_*.
\end{aligned}
\end{equation}
In the formulated minimization problem, the data fidelity and temporal penalty terms are convex and smooth, while the nuclear norm penalty term is convex but non-smooth. Accordingly, the minimization problem can be solved by use of a proximal gradient descent (PGD) method \cite{tibshirani2010proximal, combettes2011proximal}.



The convergence speed of the PGD method can be improved with momentum schemes \cite{beck2009fast, nesterov2013gradient}. To further improve the convergence speed, especially in the early iterations, an ordered subsets (OS) approach\cite{hudson1994accelerated} can be incorporated with momentum schemes\cite{kim2014combining, haase2020improved}, despite the lack of theoretical guarantees of convergence. In addition to acceleration, applying the OS approach to find an approximate solution to Eqn. \eqref{lrme-stir-opt-by-frame} offers several other significant benefits. Notably, it substantially reduces memory requirements by a factor proportional to the number $M$ of subset used. 
While the gradient with respect to all imaging frames requires $\mathcal{O}(NK)$ memory usage, the gradient corresponding to each subset only requires $\mathcal{O}(NK/M)$ storage. Furthermore, it preserve the low-rank structure of the reconstructed object function estimate at each step of the proximal gradient iteration.


For the optimization problem given by Eqn. \eqref{lrme-stir-opt}, the OS-based cost function can be expressed as\cite{kim2014combining, haase2020improved}:
\begin{equation}
\text{\textbf{J}}_{\mathcal{K}_j}(\boldsymbol{F}) = M\sum_{k \in \mathcal{K}_j}\text{\textbf{L}$_k$}(\boldsymbol{F}) + M \gamma\sum_{k \in \mathcal{K}_n}\text{\textbf{R$^{t}_k$}}(\boldsymbol{F}) + \lambda\text{\textbf{R$^{nn}$}}(\boldsymbol{F}),
\end{equation}
where, $\mathcal{K}_j$ represents the set of frame indices in the $j$-th ordered subset. The OS approach relies on the ``subset balance'' approximation\cite{kim2014combining, haase2020improved}, implying that $\textbf{J}_{\mathcal{K}_j}(\boldsymbol{F})  \approx \textbf{J}(\boldsymbol{F})$. The update procedure in PGD with OS consists of two main steps. Initially, a gradient descent step is performed moving in the negative gradient of the smooth components of the OS-based cost function, which yields
\begin{equation}\label{gradient_descent}
\begin{aligned}
\boldsymbol{F}_{j+\frac{1}{2}} &= \boldsymbol{F}_{j} - \eta \nabla \Bigl( M \sum_{k \in \mathcal{K}_j} \text{\textbf{L}$_k$}(\boldsymbol{F}_j) + M\gamma \sum_{k \in \mathcal{K}_j} \text{\textbf{R$^{t}_k$}}(\boldsymbol{F}_j) \Bigr)\\
&= \boldsymbol{F}_{j} - \eta M \sum_{ k \in \mathcal{K}_j} \Bigl( \bigl(\boldsymbol{H_{k}}^T(\boldsymbol{H_{k}}\boldsymbol{f_{k}}-\boldsymbol{g_{k}})\bigr) \otimes \boldsymbol{e}_k  + \gamma \boldsymbol{F}_j \bigl(\boldsymbol{d}_k \otimes \boldsymbol{d}_k\bigr)\Bigr).\\
\end{aligned}
\end{equation}
Above, $\eta$ is the step size and $\boldsymbol{F}_{j}$ denotes the spatiotemporal object estimate at the beginning of the $j$-th update. The first component of the gradient, associated with the data fidelity term for the $k$-th frame, results in an outer product that produces a rank-$1$ matrix. Similarly, the second component of the gradient, related to the temporal regularization term for the $k$-th frame, also yields a rank-$1$ matrix. 
Thus, the maximum rank of the gradient of the OS-based cost function $\text{\textbf{J}}_{\mathcal{K}_j}$ is bounded by $2|\mathcal{K}_j|$, where $|\mathcal{K}_j|$ denotes the number of imaging frames in the ordered subset. 





Following the gradient descent step, the proximal step is executed, where the proximal operator is applied to account for the nonsmooth component of the objective function. The proximal step corresponds to the solution of the following minimization problem:
\begin{equation}\label{prox_op}
\boldsymbol{F}_{j+1} = \operatorname{prox}_{\eta\lambda\|\cdot\|_*}\left( \boldsymbol{F}_{j+\frac{1}{2}} \right) \coloneqq \operatornamewithlimits{argmin}\limits_{\boldsymbol{F} \in \mathbb{R}^{N \times K}_{R_{\max}} } \frac{1}{2}\lVert\boldsymbol{F}-\boldsymbol{F}_{j+\frac{1}{2}}\rVert^2_2 + \eta\lambda\lVert\boldsymbol{F}\rVert_*,
\end{equation}
whose solution can be efficiently implemented via a truncated SVD factorization of $\boldsymbol{F}_{j+\frac{1}{2}}$ and the application of the soft-thresholding operator, $\mathcal{S}(.)$, to its singular values $\{\sigma_i\}$.\cite{cai2010singular} The solution of the Eqn. \eqref{prox_op} is expressed as following:
\begin{equation}
F_{j+1} = \boldsymbol{U}_{j+\frac{1}{2}}\mathcal{S}_{\eta\lambda}\left(\boldsymbol{{\Sigma}}_{j+1/2}\right)\boldsymbol{V}_{j+\frac{1}{2}}^T = \boldsymbol{U}_{j+\frac{1}{2}}\boldsymbol{\tilde{\Sigma}}_{j+\frac{1}{2}}\boldsymbol{V}_{j+\frac{1}{2}}^T,
\end{equation}
where $\boldsymbol{U}_{j+\frac{1}{2}}$,$\boldsymbol{\Sigma}_{j+\frac{1}{2}}$,$\boldsymbol{V}_{j+\frac{1}{2}}$ stem from the truncated SVD of $\boldsymbol{F}_{j+\frac{1}{2}}$ with maximum rank $R_{\max}$ and $\boldsymbol{\tilde{\Sigma}}_{j+\frac{1}{2}} \coloneqq \mathcal{S}_{\eta\lambda}\left(\boldsymbol{{\Sigma}}_{j+1/2}\right)$. 
The soft-thresholding operator, $\mathcal{S}(.)$, is defined (component-wise) as below:
\begin{equation}
  \mathcal{S}_{\eta\lambda}(\sigma_i) = \tilde{\sigma_i} =
    \begin{cases}
      \sigma_i -  \eta\lambda& \text{if } \sigma_i > \eta\lambda\\
      0, & \text{if } \sigma_i \leq \eta\lambda.
    \end{cases}       
\end{equation}
The soft-thresholding is computationally efficient and enforces a low-rank structure effectively attenuating the singular values.

\begin{algorithm}[tbh]
\caption{Low-rank matrix estimation-based spatiotemporal image reconstruction }
\label{algo:low-rank-recon}
\begin{algorithmic}
\State \textbf{Input:} $R_{\max}$, $\epsilon$, $\gamma$, $\lambda$, $\eta$, $M$
\State \textbf{Output:} $\boldsymbol{F}$
\State \textbf{Initialization:} $\boldsymbol{F}^{(0)}_1 = \boldsymbol{0}$, $t_1^{(0)} = 1$, $b= \lceil K / M \rceil$, $i = 0$ 
\While{$||\boldsymbol{F}^{(i)} - \boldsymbol{F}^{(i-1)}||_F^2/\max_{ l \leq i}||\boldsymbol{F}^{(l)} - \boldsymbol{F}^{(l-1)}||_F^2 \leq \epsilon$} 
    \State \hfill \Comment{Convergence check}
    \State $\mathcal{K} = \text{shuffle}(1, 2, ..., K)$
    \State \hfill \Comment{Randomly shuffle frames}
    \For{$j = 1$ \textbf{to} $M$}
        \State $\mathcal{K}_j = \mathcal{K}[(j-1)b + 1 : \min(jb, K)]$ 
        \State \hfill \Comment{Select the current subset of frames}
        \State \textbf{Compute gradient descent for the current subset of frames:}
        \State $\boldsymbol{F}^{(i)}_{j+\frac{1}{2}} = \boldsymbol{\bar{F}}^{(i)}_{j} - \eta \nabla \Bigl( M \sum_{k \in \mathcal{K}_j} \text{\textbf{L}$_k$}(\boldsymbol{\bar{F}}_j^{(i)}) + M\gamma \sum_{k \in \mathcal{K}_j} \text{\textbf{R$^{t}_k$}}(\boldsymbol{\bar{F}}_j^{(i)}) \Bigr)$ 
        \State\hfill \Comment{$\text{rank}(\boldsymbol{F}^{(i)}_{j+\frac{1}{2}}) \leq 2R_{max}+2b$}
        \State \textbf{Perform proximal operator:\cite{cai2010singular}}
        \State $\boldsymbol{U}_{j+\frac{1}{2}}^{(i)}, \boldsymbol{\Sigma}_{j+\frac{1}{2}}^{(i)}, \boldsymbol{V}_{j+\frac{1}{2}}^{(i)} = \text{SVD}(\boldsymbol{F}_{j+\frac{1}{2}}^{(i)}, R_{max})$ \State \hfill \Comment{Compute the truncated SVD with maximum rank $R_{max}$}
        \State $\boldsymbol{\tilde{\Sigma}}_{j+\frac{1}{2}}^{(i)} = \mathcal{S}_{\eta\gamma}(\boldsymbol{\Sigma}_{j+\frac{1}{2}}^{(i)})$ 
        \State\hfill \Comment{Soft thresholding for nuclear norm regularization}
        \State $\boldsymbol{F}^{(i)}_{j+1} = \boldsymbol{U}_{j+\frac{1}{2}}^{(i)}\boldsymbol{\tilde{\Sigma}}_{j+\frac{1}{2}}^{(i)}(\boldsymbol{V}_{j+\frac{1}{2}}^{(i)})^T$
        \State \hfill \Comment{$\text{rank}(\boldsymbol{F}^{(i)}_{j+1}) \leq R_{max}$}
        \State \textbf{Implement FISTA acceleration\cite{beck2009fast}:}
        \State $t_{j+1}^{(i)} = \frac{1 + \sqrt{1 + 4(t_{j}^{(i)})
^2}}{2}$ 
        \State \hfill \Comment{Update momentum term}
        \State $\boldsymbol{\bar{F}}_{j+1}^{(i)} = \boldsymbol{F}_{j+1}^{(i)} + \frac{t_{j}^{(i)} - 1}{t_{j}^{(i)}}(\boldsymbol{F}_{j+1}^{(i)} - \boldsymbol{F}_{j}^{(i)})$ 
        \State \hfill \Comment{FISTA update for the current iteration; $\text{rank}(\boldsymbol{\bar{F}}_{j+1}^{(i)}) \leq 2R_{max}$}
    \EndFor
    \State $\boldsymbol{F}^{(i+1)} = \boldsymbol{F}_{N+1}^{(i)}$
    \State $\boldsymbol{\bar{F}}_{1}^{(i+1)} = \boldsymbol{\bar{F}}_{N+1}^{(i)}$
    \State $t_{1}^{(i+1)} = t_{N+1}^{(i)}$
    \State $i = i + 1$ 
    \State \hfill \Comment{Increment outer loop iteration}
\EndWhile
\end{algorithmic}
\end{algorithm}


Algorithm \ref{algo:low-rank-recon} summarizes the proposed  accelerated proximal gradient descent algorithm, integrating both momentum and ordered subset techniques to efficiently find an approximate solution to the minimization problem in Eqn. \eqref{lrme-stir-opt}. The algorithm takes the following parameters as input: the maximum allowed rank $R_{\max}$; the threshold $\epsilon$ for the stopping criterion; the regularization parameter $\gamma$ for temporal regularization; the regularization parameter $\lambda$ for nuclear-norm regularization; the step size $\eta$; and the number $M$ of subsets. At the beginning of each iteration, the sequence of frame indices undergoes random shuffling. Within the subsequent inner loop, these shuffled indices are partitioned into $M$ subsets. The gradient pertaining to both the data fidelity and temporal regularization components is computed for these subset frame indices, and the gradient descent step is executed. Then, the proximal mapping associated with nuclear norm regularization is evaluated efficiently by using randomized singular value decomposition (SVD)\cite{halko2011finding} and soft thresholding. Subsequent to this step, the FISTA \cite{beck2009fast} acceleration is deployed to enhance the convergence rate. The algorithm terminates when the squared Frobenius norm of the difference between two successive iterations $||\boldsymbol{F}^{(i)} - \boldsymbol{F}^{(i-1)}||_F^2$, normalized by  its maximum $\max_{ l \leq i}||\boldsymbol{F}^{(l)} - \boldsymbol{F}^{(l-1)}||_F^2\bigr)$ over the previous iterations,  falls below the threshold $\epsilon$ defined by the user. This metric is equivalent to monitoring the norm of the gradient in smooth optimization.\cite{combettes2011proximal}

\section{Study Description}\label{study_description}

\subsection{3D PACT Imaging System Specifications for Experimental and Numerical Studies}
\begin{figure}[tbh]
\centering
\includegraphics[width=0.9\textwidth]{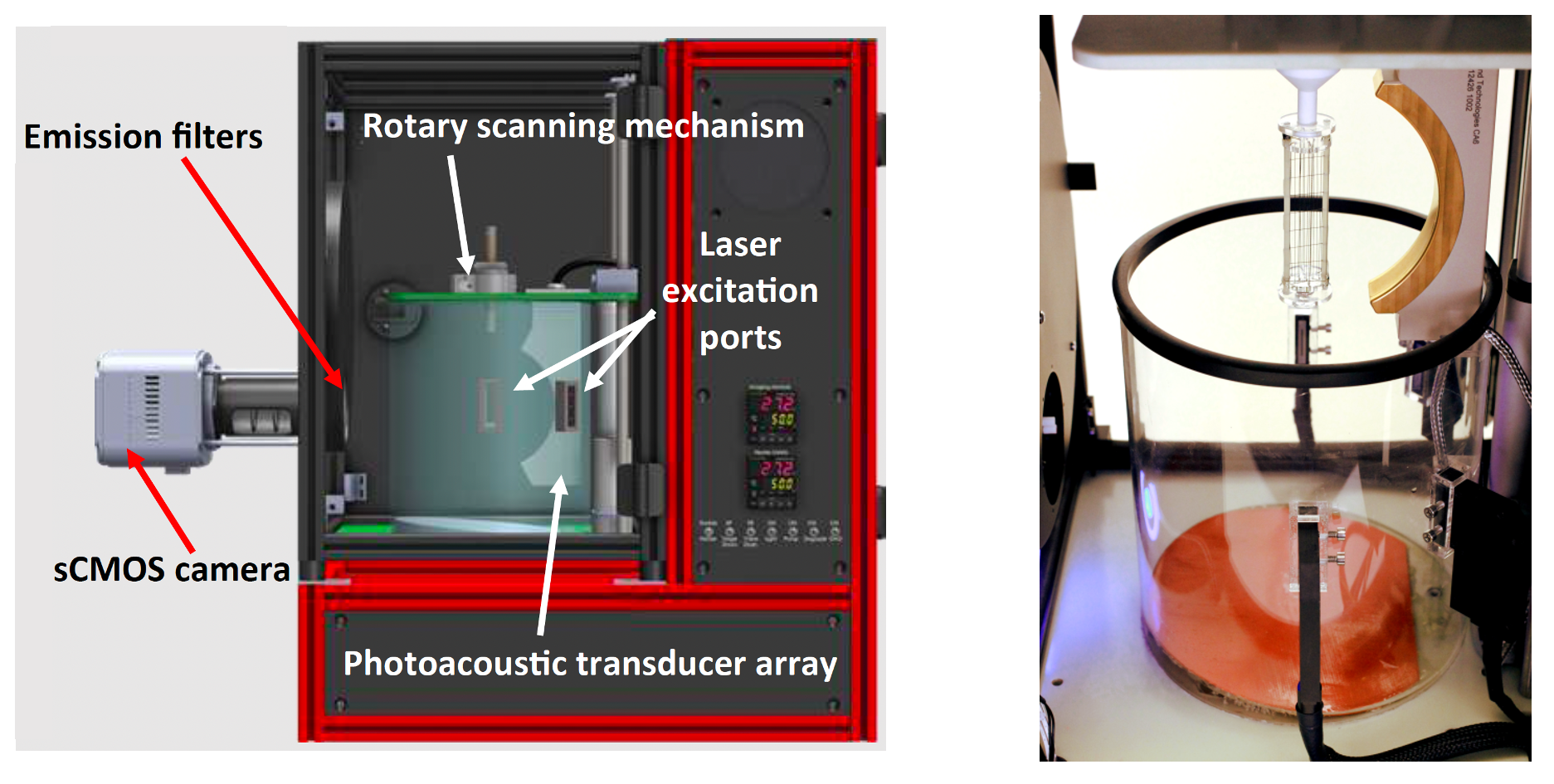}
\caption{Illustration of the TriTom imaging system (left) and an image of the imaging chamber (right).}
\label{tritom_illustration}
\end{figure}

The TriTom preclinical imaging system\cite{tritom,thompson2023characterizing} developed by PhotoSound was employed for  the experimental studies and emulated for the numerical in-silico studies. It integrates photoacoustic (PA) and fluorescence (FL) imaging modalities, harnessing their individual strengths. The system comprises a central rotary scanning stage, an optical excitation setup for PA and FL imaging, a curvilinear 96-element PA transducer array, and a fluorescence-enabled scientific complementary metal-oxide-semiconductor (sCMOS) camera. The configuration of the TriTom setup is depicted in the left panel of Figure \ref{tritom_illustration}, and the imaging chamber is shown in the right panel.


During imaging, the object is immersed in water and continuously rotated while optically stimulated by a short laser pulse with a 10Hz repetition rate. The maximum rotation speed is 10 degrees per second, and therefore completing a full 360-degree scan requires 36 seconds. Four optical fiber bundles are located on the outer circumference of the cylindrical imaging chamber, perpendicular to the scanning plane of the PA array. The laser emits pulses in the 670 to 1064 nm wavelength range at a frequency of 10 Hz, with each pulse lasting 5 ns. Acoustic waves generated by the laser excitation are detected by the PA transducer array, which comprises piezoelectric transducer elements, each measuring 1.3$\times$1.3 mm$^2$. The center frequency of the transducer elements is 6 MHz $\pm$ 10$\%$ (at -6 dB) with bandwidth $\geq$ 55$\%$. The array is vertically oriented and cylindrically focused, with the central element positioned 65 mm from the center of the imaging chamber. For fluorescence imaging, an sCMOS camera with a 2048$\times$2040 pixel resolution and a 40$\times$40 mm$^2$ field-of-view is placed outside the imaging chamber.

\begin{figure}[tbh]
\centering
\includegraphics[width=0.9\textwidth]{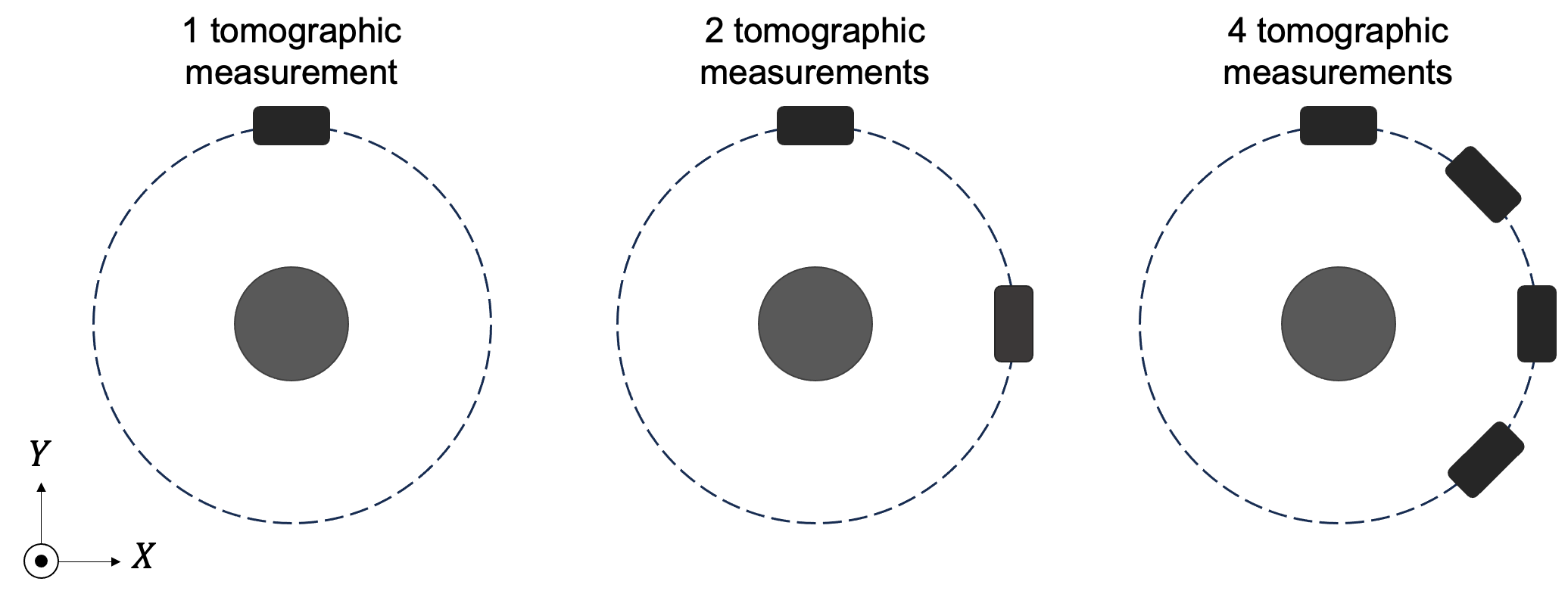}
\caption{Top view of the virtual imaging systems collecting 1, 2, or 4 tomographic views per imaging frame.} 
\label{multiple_meas}
\end{figure}

In the acoustic modeling of the TriTom imaging system, the transducer array was assumed to consist of idealized point-like transducers positioned at the central locations of the transducer elements. The acoustic simulation was implemented with a GPU-accelerated D-D imaging model\cite{wang2013accelerating} assuming acoustically homogeneous medium. Although the TriTom system has a single transducer arc, other existing 3D PACT designs (e.g. \cite{schoustra2019twente, lin2021high}) feature multiple transducer arcs, thus the numerical studies also explored scenarios where multiple tomographic measurements were acquired per imaging frame. Specifically, measurement configurations in which 2 transducer arcs separated by 90$^\circ$ and 4 transducer arcs separated by 45$^\circ$ were considered, as illustrated in Figure \ref{multiple_meas}. The sampling rate for both the experimental and numerical studies was set to 31.25 MHz, with 2048 temporal samples collected per imaging frame.


\subsection{Inverse Crime Validation Study}
\begin{figure}[tbh]
\centering
\includegraphics[width=0.8\linewidth]{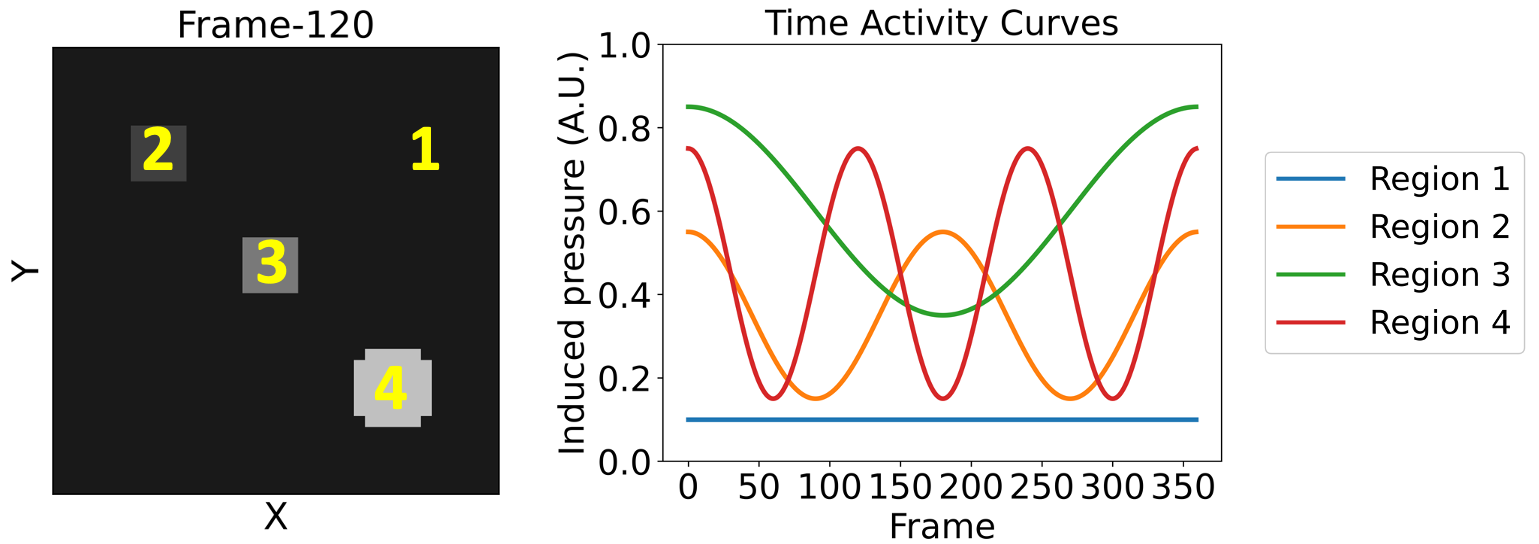}
\caption{Central z-slice of the simple rank-4 phantom at the 120-th object frame (left), time activity curves at the numbered regions in the z-slice (right). Region 1 denotes a static background.}
\label{validation_phan_ill} 
\end{figure}

To verify that Algorithm \ref{algo:low-rank-recon} was correctly implemented, an inverse crime validation study was conducted in silico in which a simple rank-4 dynamic phantom was employed. The phantom consisted of 40$\times$40$\times$3 spatial voxels and 360 object frames, with a voxel size of 0.4$\times$0.4$\times$0.4 mm$^3$. The induced pressure in the phantom was assumed constant along the z-axis and piecewise constant within the xy-plane at each object frame. The phantom was structured into four distinct regions, each characterized by varying temporal activities as shown in Figure \ref{validation_phan_ill}. The left panel of Figure \ref{validation_phan_ill} illustrates the central z-slice of the phantom at the 120-th object frame, while the right panel displays the time activity curves corresponding to the numbered regions. For each imaging frame (360 in total), 4 tomographic measurements separated by 45$^{\circ}$ were considered as depicted in the right panel of Figure \ref{multiple_meas}. Simulated acoustic pressure data were generated using a grid voxel size of 0.4$\times$0.4$\times$0.4 mm$^3$, assuming the phantom was centered within the imaging system \cite{wang2013accelerating}. The speed of sound was assumed constant at 1495 m/s, and no noise was added to the simulated measurement data.

During image reconstruction, the same computational grid that was used for generating the measurement data was employed. In the algorithm, the maximum allowed rank, $R_{max}$, was set to 4, and no temporal or nuclear norm penalties were applied ($\lambda=\gamma=0$). The step-size, $\eta$, was tuned empirically to ensure convergence. To explore the impact of the number of ordered subsets used in the randomized evaluation of the data fidelity term, three different number of ordered subsets were investigated: $M \in \{1, 2, 6\}$.

\subsection{Numerical Phantom Study}
\begin{figure}[tbh]
\centering
\includegraphics[width=0.95\linewidth]{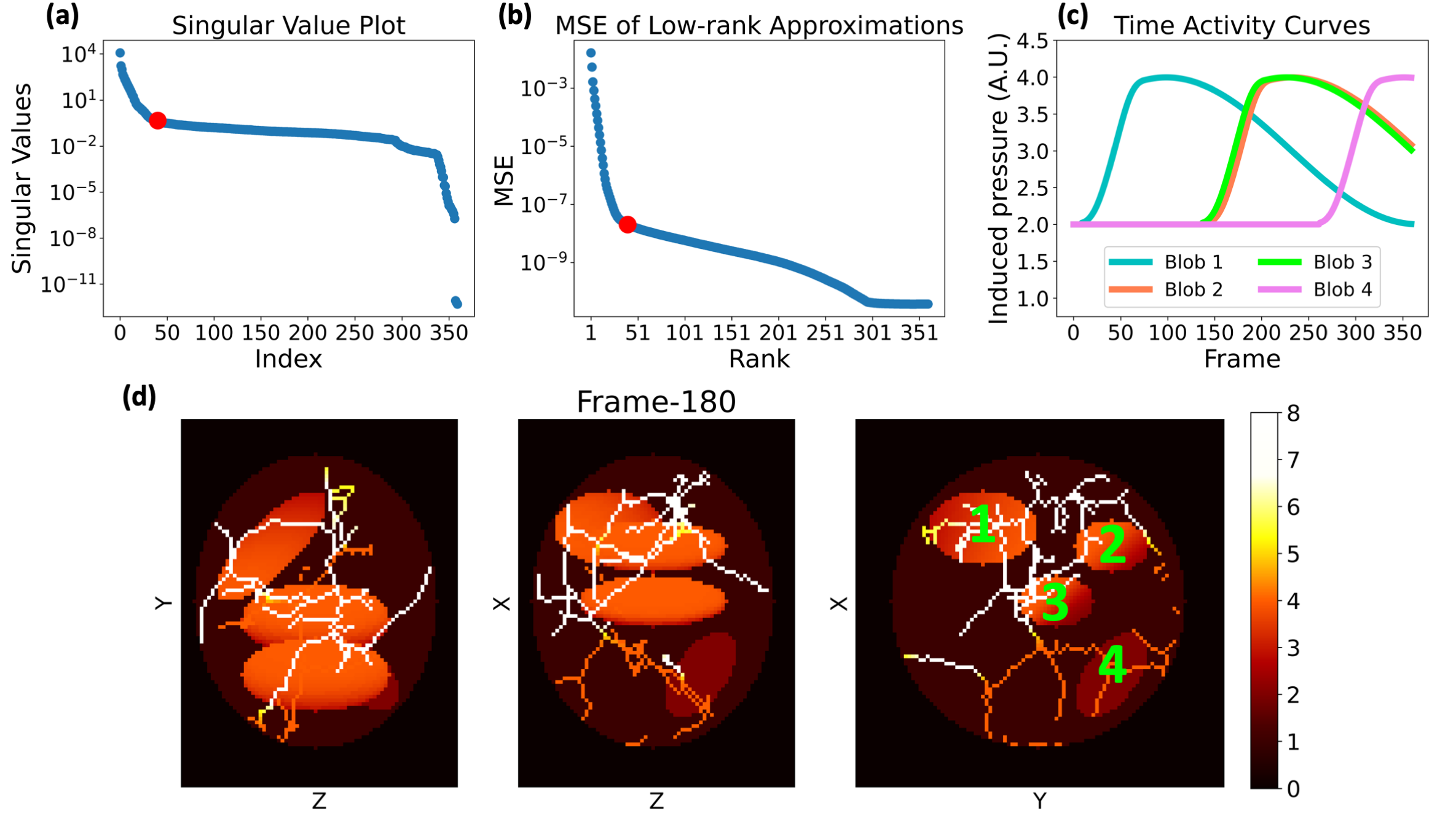}
\caption{\textbf{(a)} Singular value plot of the dynamic numerical phantom, \textbf{(b)} MSE of low-rank approximations, \textbf{(c)} time activity curves at the center of each ellipsoidal blob, \textbf{(d)} and maximum intensity projection (MIP) images of the 180-th object frame of the dynamic phantom (bottom). The ellipsoidal blobs are numbered in the MIP along the z-axis image. The red dot in panels (a) and (b) denotes the index $R_{\max} = 40$ used as maximum rank constraint in the reconstruction studies.}
\label{phantom_complex} 
\end{figure}

A dynamic numerical phantom was utilized to evaluate the performance of the proposed method through in silico experiments. The phantom consisted of 360 object frames and contained four convex ellipsoidal blobs within a larger ellipsoidal blob, along with a vasculature mimicking structure as shown in the bottom panel of Figure \ref{phantom_complex}. The size of the numerical phantom was 40$\times$40$\times$30 mm$^3$, with a voxel size of 0.4$\times$0.4$\times$0.4 mm$^3$. The time activity at each voxel was designed to mimic a contrast agent's flow along the paths from ellipsoidal blobs 1 to 2 and 3 to 4. The top right panel of Figure \ref{phantom_complex} illustrates the time activity at the center of each ellipsoidal blob, and the blobs are numbered in the bottom right panel. The singular values of the dynamic numerical phantom are shown in Figure \ref{phantom_complex}, where a rapid singular value decay is observed. The rapid singular decay indicates that the phantom can be accurately approximated with a low-rank representation, as demonstrated by the mean squared error (MSE) vs. rank plot depicted in the top middle panel of Figure \ref{phantom_complex}.




To generate the synthetic measurement data, the grid voxel size was set to 0.2$\times$0.2$\times$0.2 mm$^3$, and 360 imaging frames (corresponding to a complete rotation of the system) were simulated. The speed of sound was assumed to be constant at 1495 m/s. Zero-mean Gaussian noise with a standard deviation equivalent to a specified percentage of the maximum value of the simulated data (1\%, 3\%, or 5\%) was added to the pressure signals.

The primary objective of this numerical study was to investigate the effect of the following physical factors on the proposed method:
\begin{itemize}
\item \textbf{Number of tomographic measurements per imaging frame:} In this study, the noise level was fixed at 1\%, and three scenarios were examined with 1, 2, and 4 tomographic measurements per imaging frame, as illustrated in Figure \ref{multiple_meas}.
\item \textbf{Measurement noise level:} In this study, the number of tomographic measurements per imaging frame was set to 2, and three different noise levels were considered: 1\%, 3\%, and 5\%.
\end{itemize} 


For image reconstruction, the maximum allowed rank in the algorithm was set to $R_{max}=40$, as it allows for an accurate approximation of the dynamic numerical phantom with MSE around $10^{-8}$, as shown in Figure \ref{phantom_complex}. To avoid the discretization inverse crime, a coarser grid with voxel size 0.4$\times$0.4$\times$0.4 mm$^3$ was used for the reconstruction. The speed of sound value was the same used for simulating the data.  The threshold $\epsilon$ for the stopping criterion of Algorithm \ref{algo:low-rank-recon} was set to $2.5\times10^{-1}$. The step-size, $\eta$, was tuned empirically to ensure convergence. The number of subsets, $M$, was set to 18. To select appropriate regularization parameters, the balancing principle \cite{ito2014multi} was employed to reduce the number of tunable regularization parameters from two to one. The balancing principle rescales each regularization term by an estimate of their value at the object function $\boldsymbol{F^{true}}$. While an iterative procedure is proposed in \cite{ito2014multi} to estimate such values, for simplicity, this work assumes direct knowledge of the actual nuclear norm and the Frobenius norm of the temporal difference of $\boldsymbol{F^{true}}$. Specifically, in this study, the  solution to the dynamic image reconstruction problem was defined as: 
\begin{equation}\label{numerical_study_regularization}
\hat{\boldsymbol{F}} = \operatornamewithlimits{argmin}\limits_{\boldsymbol{F}}\frac{1}{2}\sum_{k=1}^{K}\lVert\boldsymbol{H_{k}}\boldsymbol{f_{k}}-\boldsymbol{g_{k}}\rVert^2 + \kappa \left(\frac{\sum_{k=1}^{K-1}\lVert\boldsymbol{f_{k+1}}-\boldsymbol{f_{k}}\rVert^2}{\sum_{k=1}^{K-1}\lVert\boldsymbol{f_{k+1}^{true}}-\boldsymbol{f_{k}^{true}}\rVert^2} + \frac{\lVert\boldsymbol{F}\rVert_*}{\lVert\boldsymbol{F^{true}}\rVert_*}\right).
\end{equation}
 For each case, four different values of the parameter $\kappa$ were explored: $\kappa \in \{10^{-4}\lVert \boldsymbol{G} \rVert_F^2, 5\times10^{-4}\lVert \boldsymbol{G} \rVert_F^2, 2.5\times10^{-3}\lVert \boldsymbol{G} \rVert_F^2, 1.25\times10^{-2}\lVert \boldsymbol{G} \rVert_F^2\}$. 
 The $\kappa$ parameter yielding the best average normalized squared error (nSE) over frames was selected \cite{pereverzev2005adaptive}. The nSE for each frame was computed as:
\begin{equation}
\text{nSE} = ||\boldsymbol{f_k^{true}} -\hat{\boldsymbol{f_k}}||_2^2/\operatornamewithlimits{max}\limits_{\boldsymbol{k}}||\boldsymbol{f_k^{true}}||_2^2.
\label{norm_se}
\end{equation}
Specifically, the regularization parameter value that yielded the optimal results was $\kappa=5\times10^{-4}\lVert \boldsymbol{G} \rVert_F^2$ when 4 tomographic measurements per frame were available, and $\kappa=2.5\times10^{-3}\lVert \boldsymbol{G} \rVert_F^2$ for all other cases. 




\subsection{Experimental Study}
\begin{figure}[tbh]
\centering
\includegraphics[width=0.85\textwidth]{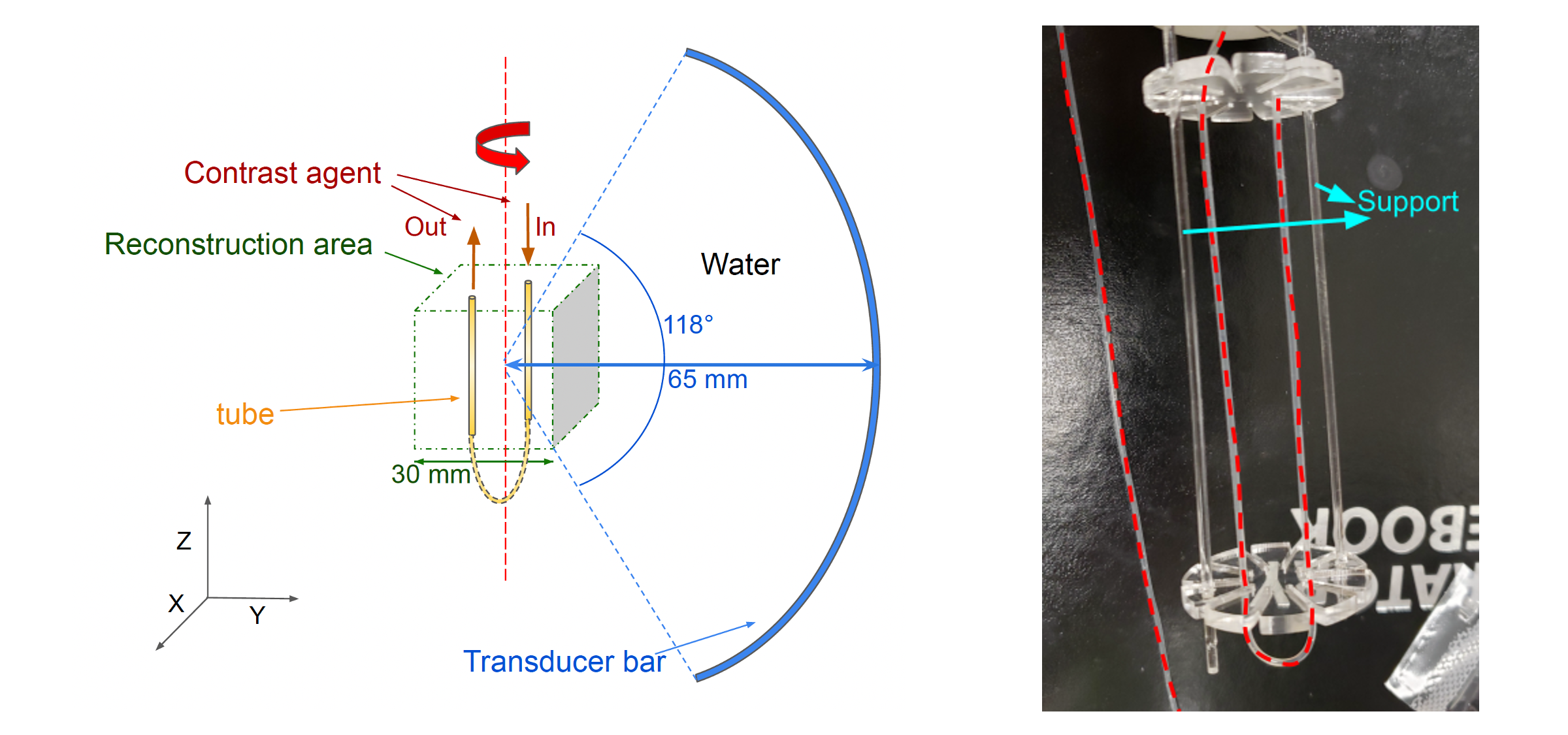}
\caption{An illustration (left) and a picture (right) of the experimental dynamic phantom.} 
\label{experimental_setup_illustration}
\end{figure}

An open-ended dynamic flow phantom was constructed by bending a silicone tube into a U-shaped structure, as depicted in Figure \ref{experimental_setup_illustration}. The left panel of the figure provides an illustration of the phantom within the imaging chamber, while the right panel shows an actual photograph of the physical phantom. The inner diameter of the silicone tube was 0.0635 cm, and the outer diameter was 0.1194 cm. The dynamic phantom, positioned at the center of the imaging system, was illuminated by a laser pulse with a wavelength of 770 nm and an energy of 100 mJ (before entering the fiber optic light delivery unit). PACT data were acquired with the TriTom imaging system during the injection of a photoacoustic-fluorescent contrast agent (PAtIR\cite{patir}) through one end of the tube. Over a span of 36 seconds, the dynamic phantom underwent scanning which resulted in 360 imaging frames, with a frame rate of 0.1 seconds and a single tomographic measurement per imaging frame. Two-dimensional fluorescence images were concurrently gathered, which serve as reference for the time evolution of the contrast agent concentration within the tube.

Based on region illuminated by laser, the reconstruction volume was set to a region of 30$\times$30$\times$30 mm$^3$ located at the center of the imaging system. The voxel size was set to 0.4$\times$0.4$\times$0.4 mm$^3$, resulting in 75$\times$75$\times$75 spatial voxels. The maximum allowed rank, $R_{max}$, was set to 40, and the number of subsets, $M$, was set to 18. For the stopping criterion of Algorithm \ref{algo:low-rank-recon}, the threshold $\epsilon$ was set to $5\times10^{-1}$. The step-size, $\eta$, was tuned empirically to ensure convergence.
To calibrate the speed of sound, multiple static estimates of the dynamic object were reconstructed using the Universal Back-Projection (UBP) algorithm\cite{xu2005universal} assuming speed of sound values in the range of 1480-1520 m/s, with 5 m/s increments. The speed of sound value of 1495 m/s resulted in the best visual appearance and was selected to perform the dynamic image reconstruction. 
To choose the regularization parameters, 9 dynamic image reconstructions where performed for all possible combinations of   $\gamma \in\{5.5\times10^0, 5.5\times10^1, 5.5\times10^2\}$, and $\lambda \in\{10^{-1}, 10^{0}, 10^{1}\}$. Ultimately, the spatiotemporal object estimate that most closely captured the observed dynamic changes in the reference fluorescence images, as determined through visual examination, was selected. The corresponding regularization parameters were $\gamma=5.5\times10^1$ and $\lambda=10^{0}$.




\section{Results}\label{results}
\subsection{Inverse Crime Validation Study Results}

In this inverse crime study, the step sizes $\eta$ for $M = 1, 2, 6$ were empirically tuned to ensure convergence, resulting in values of $\{10^{-4}, 10^{-4}, 3.3\times10^{-5}\}$, respectively. The algorithm was allowed to run for 2500 iterations in each case to ensure convergence of the solution up to numerical precision. For the purpose of this validation study, Algorithm \ref{algo:low-rank-recon} was modified to re-evaluate the total data fidelity term $\text{\textbf{L}}(\boldsymbol{F}_{i})$, which accumulates the contributions from all time frames, at each iteration.

\begin{figure}[tbh]
\centering
\includegraphics[width=0.8\textwidth]{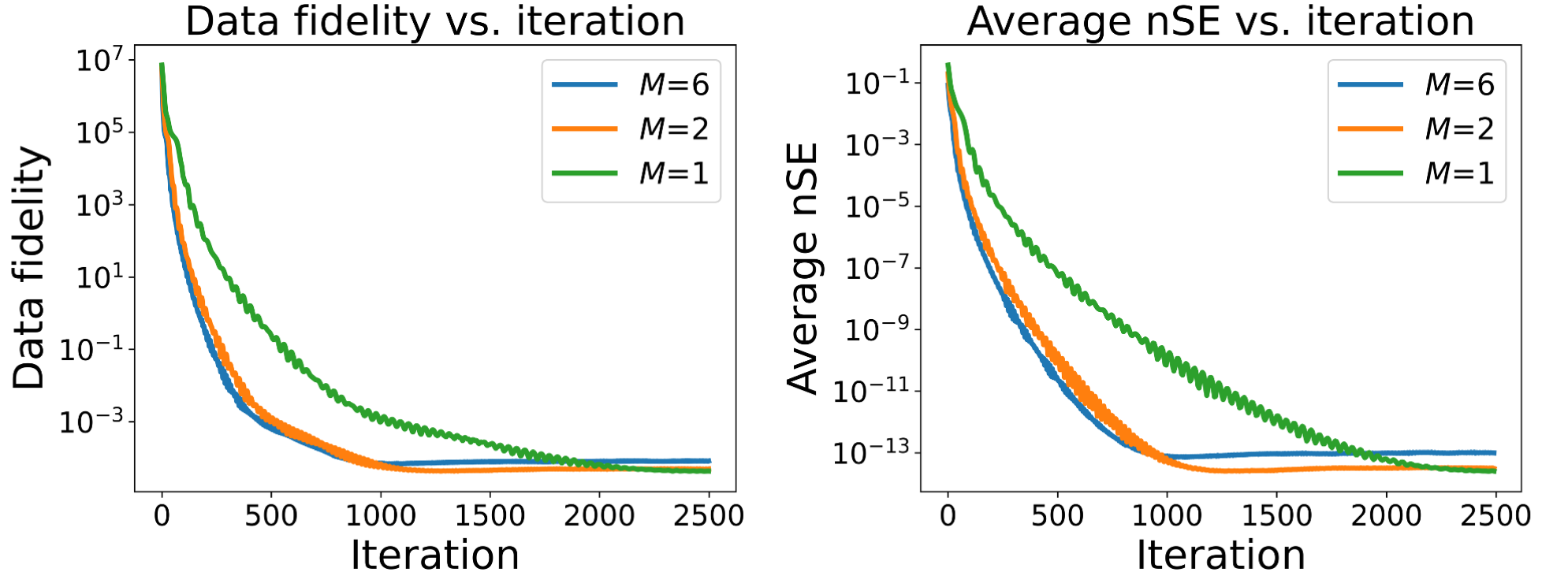}
\caption{The data fidelity vs. the iterations (left), and the average nSE vs. iterations (right). The plots confirm the correct implementation of the algorithm and illustrate how increasing the number of subsets can accelerate the reduction of the data fidelity term. Additionally, the plots illustrate the similar order-of-magnitude reduction in both data fidelity and average nSE values for every number of subsets $M\in \{1, 2, 6\}$.} 
\label{validation_study_results}
\end{figure}

The results of the validation study are depicted in Figure \ref{validation_study_results}. The left panel exhibits the data fidelity $\text{\textbf{L}}(\boldsymbol{F}_{i})$ vs. iteration count, while the right panel presents the average nSE vs. iteration count. For all values of $M$, A significant decrease of approximately 11 and 13 orders of magnitude can be observed in the data fidelity term and average nSE, respectively. This indicates that the proposed method using momentum-acceleration in combination with ordered subsets (case $M \in\{2, 6\}$), although lacking theoretical guarantee of converge, can achieve (up to machine precision) to the same object estimate produced by the momentum-accelerated proximal gradient descent without subsampling method ($M=1$).
It is also evident that a larger number of subsets improves the convergence speed, particularly in the early iterations. This also translate in possibly faster time to solution, as the cost of each iteration is dominated by the evaluation of the imaging operator, while the time spent in performing the truncated SVD factorization using randomized method is neglegible. In the numerical studies presented here, the computational time required per iteration was approximately 15 minutes on a workstation (AMD EPYC 7702P 64-Core processor, 32GB RAM, one Nvidia Geforce RTX 2080 graphic processing unit) independently of the number $M$ of subsets used. 

In summary, this validation study demonstrates the correct implementation, efficiency, and robust convergence, despite lack of theoretical guarantees, of the proposed method.

\subsection{Numerical Phantom Study Results}

\begin{figure}[tbh]
\centering
\includegraphics[width=0.99\textwidth]{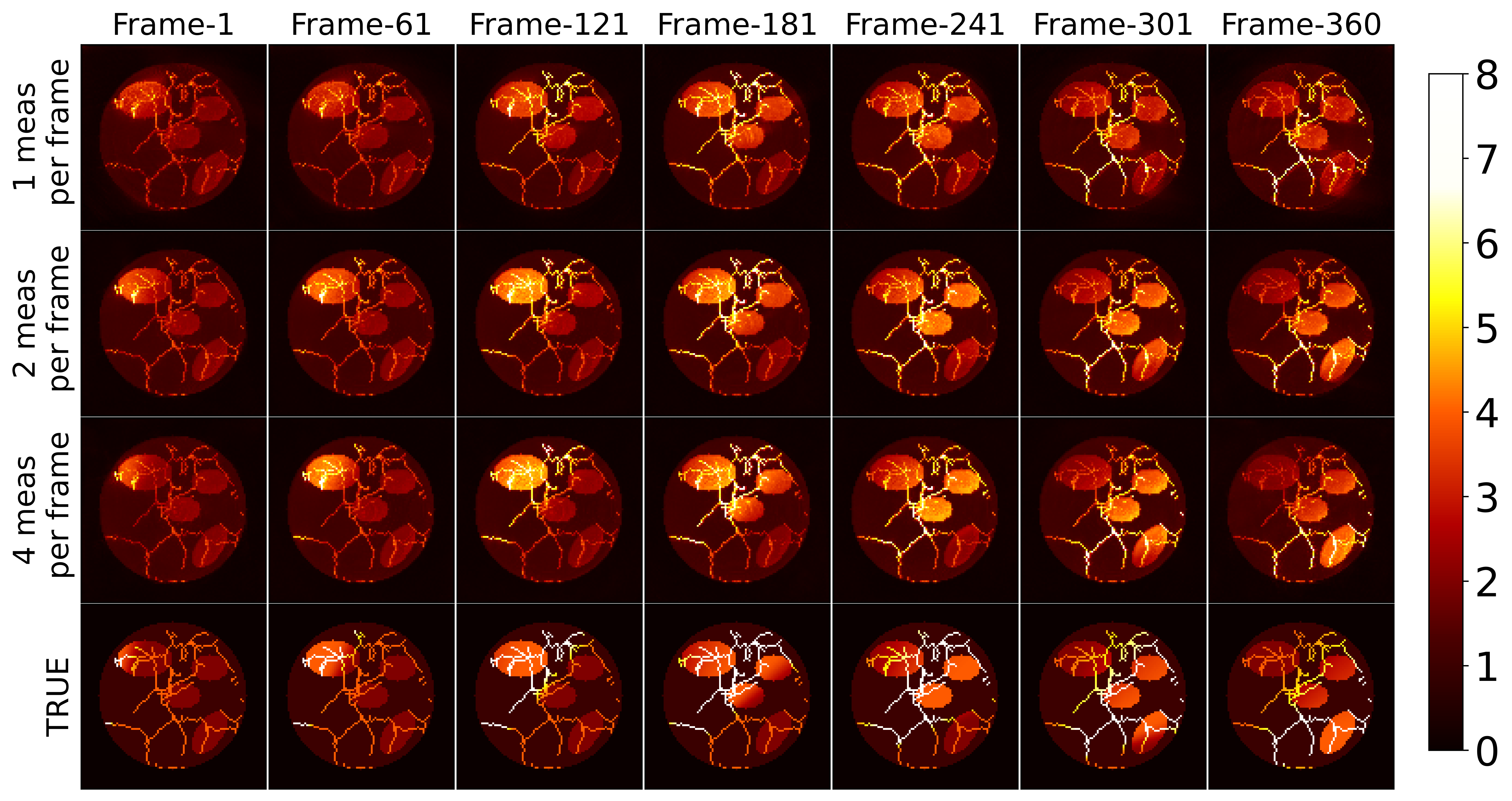}
\caption{Selected MIP images (along the z-axis) depicting the numerical phantom and corresponding spatiotemporal estimates from simulated data with varying numbers of tomographic measurements per imaging frame. A close examination reveals improved reconstruction accuracy with an increased number of tomographic measurements. Notably, the reconstructions using 2 and 4 tomographic measurements per imaging frame more accurately capture the intensity change in blob-4 (located at the bottom right) compared to the one using a single tomographic measurement per imaging frame. This observation is further supported by the TACs shown in Figure \ref{numerical_tomo_meas_tac}. A video featuring the spatiotemporal evolution of the numerical phantom and its reconstructed estimates is available as supplemental material.} 
\label{numerical_tomo_meas_mip}
\end{figure}
\begin{figure}[tbh]
\centering
\includegraphics[width=0.99\textwidth]{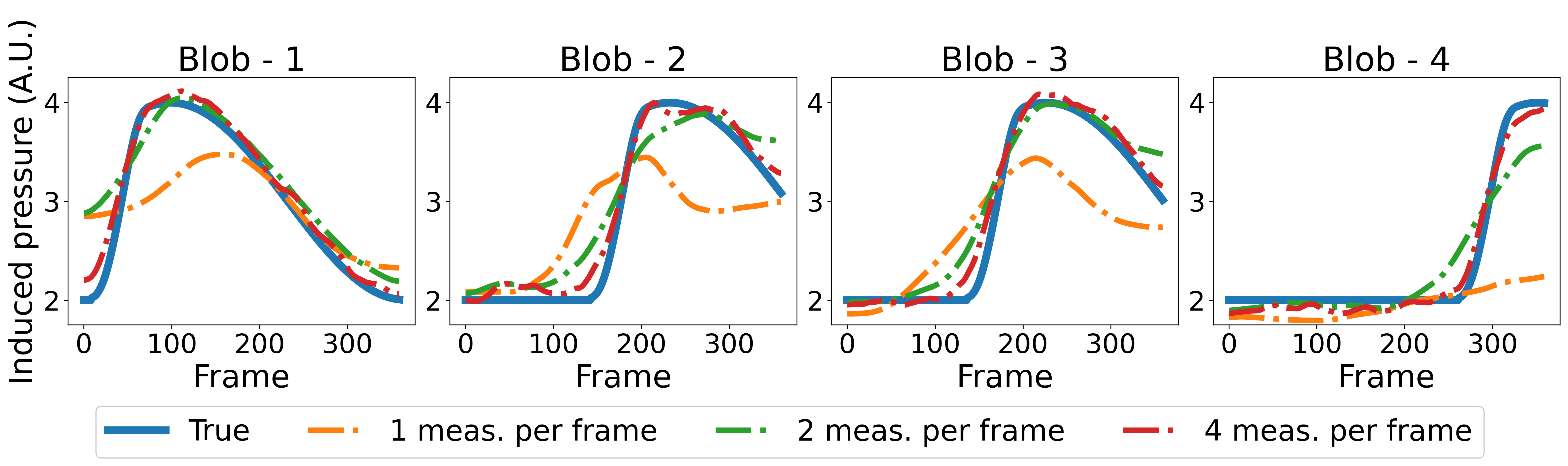}
\caption{TACs at each ellipsoidal blob's center, comparing the numerical phantom with its estimate reconstructed from different numbers of tomographic measurements per imaging frame. The noise level was kept at $1\%$. The curves demonstrate that increasing the number of measurements enhances the fidelity of temporal activities in the reconstructions.} 
\label{numerical_tomo_meas_tac}
\end{figure}
\begin{figure}[tbh]
\centering
\includegraphics[width=0.8\textwidth]{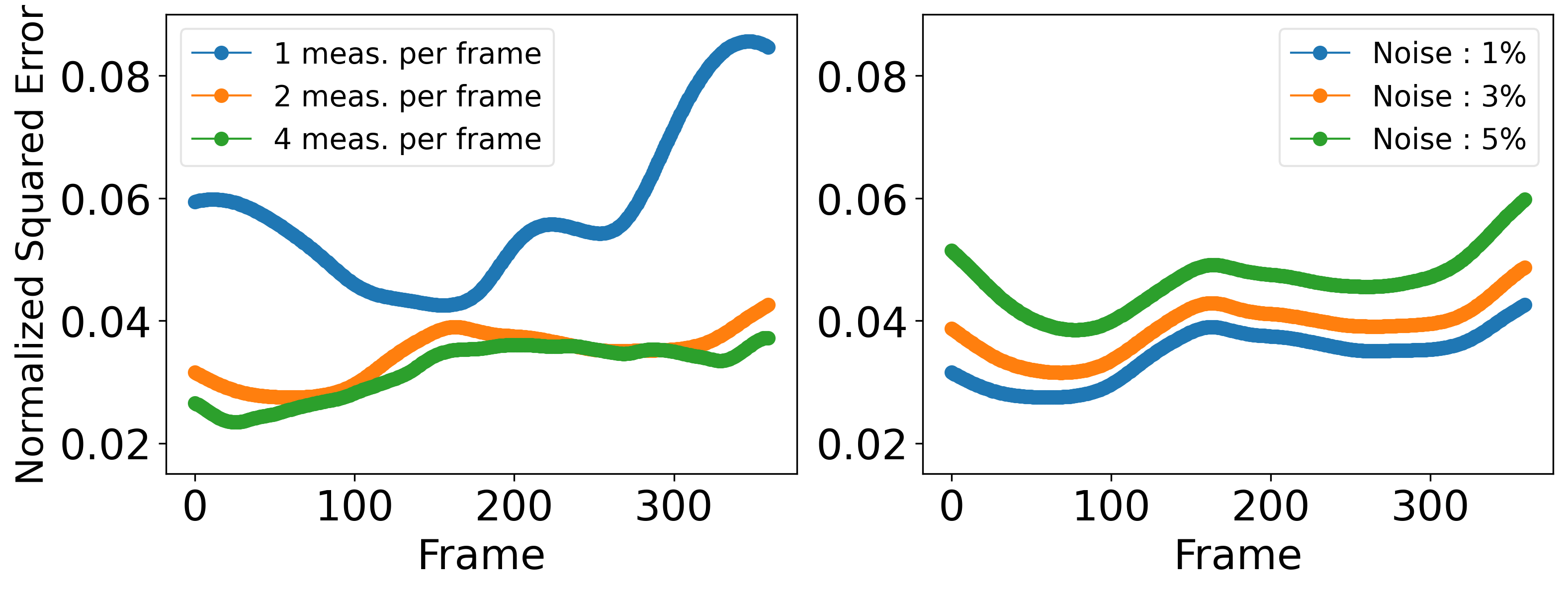}
\caption{\textbf{Left}: Normalized Squared Error (nSE) vs. the imaging frame number for reconstructions with varying numbers of tomographic measurements per imaging frame; the noise level was fixed at $1\%$. A direct correlation between an increase in measurements and reduction in nSE is observable, thus highlighting the improved reconstruction accuracy. \textbf{Right}: nSE vs.\ the imaging frame number for reconstructions from data with varying noise levels; the number of tomographic measurements per frame was fixed at 2. As anticipated, the nSE exhibits an upward trend with increasing noise levels.} 
\label{numerical_nse}
\end{figure}

\paragraph{Sensitivity to the number of tomographic measurements per imaging frame:}
Figure \ref{numerical_tomo_meas_mip} displays a selection of MIP images depicting the numerical phantom and the spatiotemporal estimates from simulated data with varying numbers of tomographic measurements per imaging frame. A video featuring the spatiotemporal evolution of the numerical phantom and its corresponding estimates is available as supplemental material\cite{video}. Upon careful inspection, it is apparent that increasing the number of tomographic measurements per imaging frame leads to more accurate estimates of the dynamic object. For instance, object estimates reconstructed from data with 2 and 4 measurements per imaging frame correctly capture the increase in intensity of blob-4 after frame-301. However, this intensity change is less prominent in the dynamic estimate reconstructed from data with only one tomographic measurement per imaging frame, which also exhibits artifacts around blob-4. In addition, the estimate reconstructed using 4 measurements per imaging frame better captures the intensity difference between the two ends of the blob-1 in frame-1. These observations are more evident in Figure \ref{numerical_tomo_meas_tac}, which shows the time-activity curves (TACs) at each ellipsoidal blob's center in both the numerical phantom and the spatiotemporal estimates from data with different numbers of tomographic measurements per imaging frame. Notably, the TACs for the reconstruction from 4 tomographic measurements per imaging frame closely aligns with those of the numerical phantom. The normalized squared error (nSE) vs. imaging frame plot for the reconstructions with varying numbers of tomographic measurements per imaging frame is depicted in the left panel of Figure \ref{numerical_nse}. Both the TACs and nSE vs. frame plots confirm the observation that a higher number of tomographic measurements per imaging frame leads to more accurate spatiotemporal reconstructions.



\begin{figure}[tbh]
\centering
\includegraphics[width=0.99\textwidth]{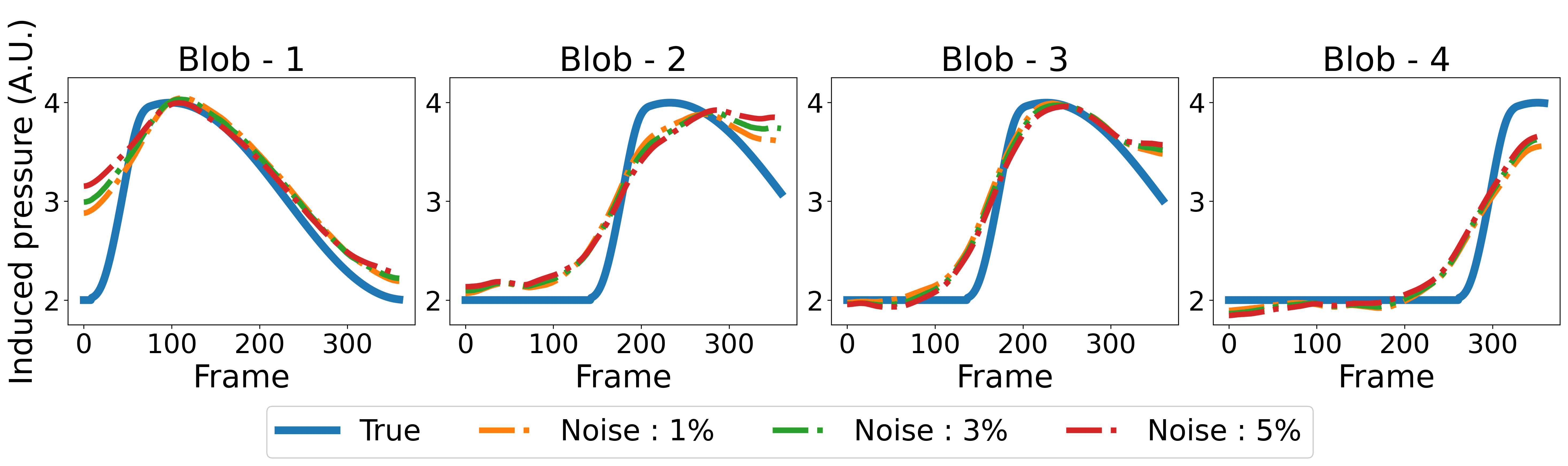}
\caption{TACs at each ellipsoidal blob's center, comparing the true phantom with reconstructions from data with different noise levels; number of tomographic measurements per frame was fixed at 2. It is seen that the recovered TACs are close to each other; which shows the robustness of the algorithm against the noise level.} 
\label{numerical_noise_tac}
\end{figure}

\paragraph{Sensitivity to measurement noise:} Figure \ref{numerical_noise_tac} displays the TACs at each ellipsoidal blob's center in both the numerical phantom and its spatiotemporal estimates from data with varying noise levels, when the number of tomographic measurements per frame is kept at two. Notably, the estimated TACs remain  similar across all noise levels, highlighting the algorithm's robustness in capturing dynamic changes even in the presence of increased noise. Figure \ref{numerical_nse} (right panel) shows the nSE vs. imaging frame plot for estimates reconstructed from data with varying noise levels. As anticipated, the nSE exhibits an upward trend with increasing noise levels.\\

The numerical phantom study results underscore the effectiveness of the proposed method in accurately estimating dynamic changes, when limited tomographic measurements per frame are available. The study reveals that the ill-posed nature of the dynamic reconstruction problem diminishes significantly as the number of tomographic measurements per imaging frame increases. Notably, the method's robustness in faithfully capturing the temporal dynamics of the object remains evident even in the presence of increased noise levels. These findings collectively underscore the algorithm's reliability and potential significance in addressing challenges associated with dynamic imaging scenarios.


\subsection{Experimental Study Results}

\begin{figure}[tbh]
\centering
\includegraphics[width=0.8\linewidth]{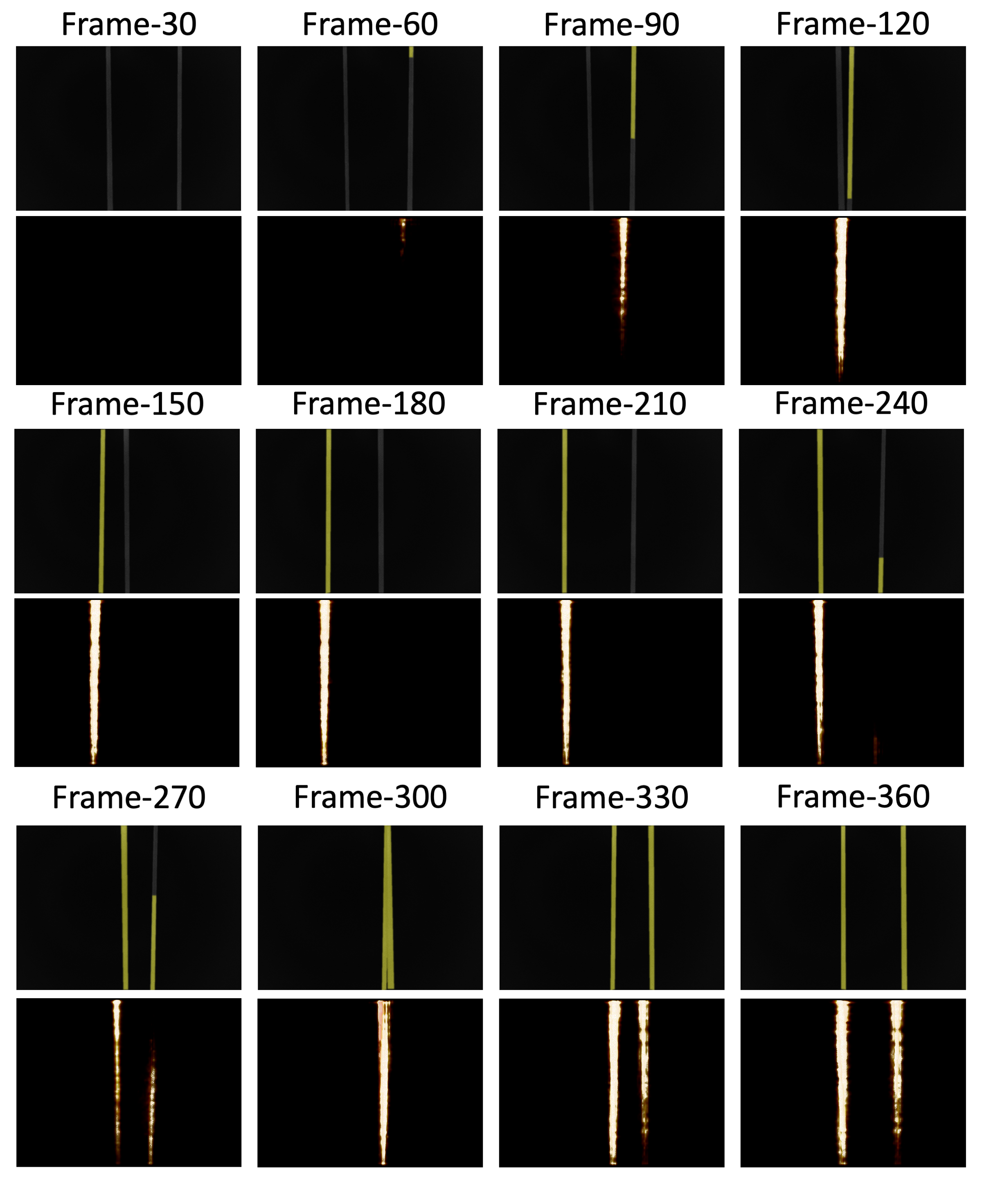}
\caption{Sample instances of the dynamic recovery from PACT data (at the bottom of each row) and reference fluorescence images (at the top of each row). The spatiotemporal evolution of the contrast agent inferred from the spatiotemporal estimates reconstructed from PACT data is in strong agreement with the reference images collected by the fluorescence-enabled sCMOS camera. A  video featuring the raw images, visually enhanced images through the segmentation of the tube and contrast agent, and ParaView visualizations of the spatiotemporal PACT object estimate, is provided in the supplemental materials.}
\label{exp_results}
\end{figure}

Figure \ref{exp_results} displays dynamic PACT images reconstructed from experimental data (bottom row), along with the corresponding reference fluorescence images (top row).  The fluorescence images were processed to suppress the background, and the contrast agent is highlighted in yellow. This enhancement was accomplished through manual segmentation of the tube and contrast agent from the raw fluorescence images. The spatiotemporal PACT object estimates were visualized using ParaView \cite{ParaView}. To ensure a qualitatively close alignment of the field-of-view between the ParaView visualization of the spatiotemporal PACT object estimate and the 2D fluorescence images, the view angle in ParaView was adjusted manually at each frame; however, a slight misalignment remains. A  video featuring the raw images, visually enhanced images through the segmentation of the tube and contrast agent, and ParaView visualizations of the spatiotemporal PACT object estimate, is provided in the supplemental materials\cite{video}.

In particular, when examining frames 30 to 120, one can readily observe the accurate recovery of dynamic contrast flow within the tube. A similar trend is evident in frames 210 to 300, with the lower right section of frame-240 showing the presence of the contrast agent, albeit with reduced contrast. This reduced contrast is also noticeable in specific frames such as 330 and 360, possibly due to light obstruction by other tube segments. Nevertheless, the overall effective recovery of dynamic flow remains apparent, as convincingly demonstrated in the supplementary video. This highlights the effectiveness and practicality of the proposed method for experimental data beyond simulated measurements, even when only one tomographic measurement per imaging frame is available.

Furthermore, it is worth noting that the spatiotemporal reconstruction offers a frame rate of 0.1 seconds. In contrast, an FBFIR technique would provide a frame rate of 36 seconds, which is the time required for a complete tomographic measurement. This underscores the reconstruction method's potential for monitoring dynamic physiological processes that demand enhanced frame rate. This can enhance the value of 3D PACT systems that involve rotating measurement gantries  for preclinical research, enabling  spatiotemporal image reconstruction from limited tomographic measurements per imaging frame.

\section{Conclusion and Discussion}\label{conclusion}

This study presents an accurate and computationally efficient LRME-STIR method for dynamic PACT attuned for commercially available volumetric imagers that employ a rotating measurement gantry in which the tomographic data are sequentially acquired. The implementation of the method was verified by an inverse-crime numerical validation study. The effect of varying number of tomographic measurements per imaging frame and the noise level on the method's accuracy was investigated in an in-silico numerical study. The experimental study demonstrated the LRME-STIR method's ability to reconstruct the flow of a contrast agent at a frame rate of 0.1 s, even when only a single tomographic measurement per imaging frame was available. Numerical and experimental studies confirm the accuracy of the proposed technique. Thus, this work will potentially have an immediate and sustained positive impact by creating a new capacity to perform dynamic 3D PACT.


The LRME-STIR method proposed in this study employs an accelerated proximal gradient descent method combined with an ordered subsets approach, distinguishing it from previously proposed LRME-STIR methods\cite{liang2007spatiotemporal, brinegar2009real, haldar2010spatiotemporal, lingala2011accelerated}. This approach yields a rapid and computationally efficient algorithm with reduced memory demands. Nonetheless, it should be noted that the proposed approach lacks theoretical convergence guarantees. While in theory the method may exhibit instabilities, numerical and experimental studies showcase the robustness of the proposed method when the regularization strength and step size are properly chosen.

Future enhancements to the method may involve exploring different heuristics to enhance its stability\cite{kim2014combining, haase2020improved}. Subsequent investigations may focus on evaluating the method's efficacy in imaging complex dynamic physiological processes and quantifying relevant parameters, such as wash-in and wash-out rates for tumor vascular perfusion \cite{hupple2018light, cam2023dynamic}. These evaluations might encompass both in-vivo and in-silico experiments, further advancing the understanding and application of the proposed method.

\section*{Disclosures}
The author S. Ermilov discloses ownership interest in Photosound Technologies, Inc. Other
authors have no relevant financial interests and no potential conflicts of interest to disclose.

\section*{Acknowledgements}
This work was supported in part by the National Institutes of Health (NIH Grant Nos. OD023029 and EB031585).

\section*{Code, Data, and Materials Availability}
Data and code are available upon request. Please contact Dr. Villa at \linkable{uvilla@oden.utexas.edu}.





\bibliography{report}   

\begin{thebibliography}{10}

\bibitem{wang2003noninvasive}
X.~Wang, Y.~Pang, G.~Ku, {\em et~al.}, ``Noninvasive laser-induced
  photoacoustic tomography for structural and functional in vivo imaging of the
  brain,'' {\em Nature biotechnology} {\bf 21}(7), 803--806  (2003).

\bibitem{brecht2009whole}
H.-P. Brecht, R.~Su, M.~Fronheiser, {\em et~al.}, ``Whole-body
  three-dimensional optoacoustic tomography system for small animals,'' {\em
  Journal of biomedical optics} {\bf 14}(6), 064007--064007  (2009).

\bibitem{yang2007functional}
S.~Yang, D.~Xing, Q.~Zhou, {\em et~al.}, ``Functional imaging of
  cerebrovascular activities in small animals using high-resolution
  photoacoustic tomography,'' {\em Medical physics} {\bf 34}(8), 3294--3301
  (2007).

\bibitem{kagadis2010vivo}
G.~C. Kagadis, G.~Loudos, K.~Katsanos, {\em et~al.}, ``In vivo small animal
  imaging: current status and future prospects,'' {\em Medical physics} {\bf
  37}(12), 6421--6442  (2010).

\bibitem{loudos2011current}
G.~Loudos, G.~C. Kagadis, and D.~Psimadas, ``Current status and future
  perspectives of in vivo small animal imaging using radiolabeled
  nanoparticles,'' {\em European journal of radiology} {\bf 78}(2), 287--295
  (2011).

\bibitem{franc2008small}
B.~L. Franc, P.~D. Acton, C.~Mari, {\em et~al.}, ``Small-animal spect and
  spect/ct: important tools for preclinical investigation,'' {\em Journal of
  nuclear medicine} {\bf 49}(10), 1651--1663  (2008).

\bibitem{carmeliet2005angiogenesis}
P.~Carmeliet, ``Angiogenesis in life, disease and medicine,'' {\em Nature} {\bf
  438}(7070), 932--936  (2005).

\bibitem{carmeliet2000angiogenesis}
P.~Carmeliet and R.~K. Jain, ``Angiogenesis in cancer and other diseases,''
  {\em Nature} {\bf 407}(6801), 249--257  (2000).

\bibitem{small_animal_pact}
J.~Xia and L.~V. Wang, ``Small-animal whole-body photoacoustic tomography: A
  review,'' {\em IEEE Transactions on Biomedical Engineering} {\bf 61}(5),
  1380--1389  (2014).

\bibitem{li2017single}
L.~Li, L.~Zhu, C.~Ma, {\em et~al.}, ``Single-impulse panoramic photoacoustic
  computed tomography of small-animal whole-body dynamics at high
  spatiotemporal resolution,'' {\em Nature biomedical engineering} {\bf 1}(5),
  0071  (2017).

\bibitem{upputuri2017dynamic}
P.~K. Upputuri and M.~Pramanik, ``Dynamic in vivo imaging of small animal brain
  using pulsed laser diode-based photoacoustic tomography system,'' {\em
  Journal of biomedical optics} {\bf 22}(9), 090501--090501  (2017).

\bibitem{shi2019thermosensitive}
S.~Shi, X.~Wen, T.~Li, {\em et~al.}, ``Thermosensitive biodegradable copper
  sulfide nanoparticles for real-time multispectral optoacoustic tomography,''
  {\em ACS applied bio materials} {\bf 2}(8), 3203--3211  (2019).

\bibitem{manohar2019current}
S.~Manohar and M.~Dantuma, ``Current and future trends in photoacoustic breast
  imaging,'' {\em Photoacoustics} {\bf 16}, 100134  (2019).

\bibitem{lin2018single}
L.~Lin, P.~Hu, J.~Shi, {\em et~al.}, ``Single-breath-hold photoacoustic
  computed tomography of the breast,'' {\em Nature communications} {\bf 9}(1),
  2352  (2018).

\bibitem{gamelin2009real}
J.~Gamelin, A.~Maurudis, A.~Aguirre, {\em et~al.}, ``A real-time photoacoustic
  tomography system for small animals,'' {\em Optics express} {\bf 17}(13),
  10489--10498  (2009).

\bibitem{he2017plasmonic}
W.~He, K.~Ai, C.~Jiang, {\em et~al.}, ``Plasmonic titanium nitride
  nanoparticles for in vivo photoacoustic tomography imaging and photothermal
  cancer therapy,'' {\em Biomaterials} {\bf 132}, 37--47  (2017).

\bibitem{upputuri2017high}
P.~K. Upputuri, V.~Periyasamy, S.~K. Kalva, {\em et~al.}, ``A high-performance
  compact photoacoustic tomography system for in vivo small-animal brain
  imaging,'' {\em JoVE (Journal of Visualized Experiments)} (124), e55811
  (2017).

\bibitem{shan2020vivo}
T.~Shan, Y.~Zhao, S.~Jiang, {\em et~al.}, ``In-vivo hemodynamic imaging of
  acute prenatal ethanol exposure in fetal brain by photoacoustic tomography,''
  {\em Journal of biophotonics} {\bf 13}(5), e201960161  (2020).

\bibitem{brecht20183d}
H.~P. Brecht, V.~Ivanov, D.~S. Dumani, {\em et~al.}, ``A 3{D} imaging system
  integrating photoacoustic and fluorescence orthogonal projections for
  anatomical, functional and molecular assessment of rodent models,'' in {\em
  Photons Plus Ultrasound: Imaging and Sensing 2018},   {\bf 10494}, 81--88,
  SPIE  (2018).

\bibitem{lin2021high}
L.~Lin, P.~Hu, X.~Tong, {\em et~al.}, ``High-speed three-dimensional
  photoacoustic computed tomography for preclinical research and clinical
  translation,'' {\em Nature communications} {\bf 12}(1), 882  (2021).

\bibitem{ermilov2009laser}
S.~A. Ermilov, T.~Khamapirad, A.~Conjusteau, {\em et~al.}, ``Laser optoacoustic
  imaging system for detection of breast cancer,'' {\em Journal of biomedical
  optics} {\bf 14}(2), 024007--024007  (2009).

\bibitem{4d_pact}
L.~Xiang, B.~Wang, L.~Ji, {\em et~al.}, ``4-{D} photoacoustic tomography,''
  {\em Scientific reports} {\bf 3}(1), 1--8  (2013).

\bibitem{cai2014cine}
J.-F. Cai, X.~Jia, H.~Gao, {\em et~al.}, ``Cine cone beam {CT} reconstruction
  using low-rank matrix factorization: algorithm and a proof-of-principle
  study,'' {\em IEEE transactions on medical imaging} {\bf 33}(8), 1581--1591
  (2014).

\bibitem{spatiotemporal_pet}
M.~Wernick, E.~Infusino, and M.~Milosevic, ``Fast spatio-temporal image
  reconstruction for dynamic {PET},'' {\em IEEE Transactions on Medical
  Imaging} {\bf 18}(3), 185--195  (1999).

\bibitem{ding2017dynamic}
Q.~Ding, M.~Burger, and X.~Zhang, ``Dynamic {SPECT} reconstruction with
  temporal edge correlation,'' {\em Inverse Problems} {\bf 34}(1), 014005
  (2017).

\bibitem{haldar2010spatiotemporal}
J.~P. Haldar and Z.-P. Liang, ``Spatiotemporal imaging with partially separable
  functions: A matrix recovery approach,'' in {\em 2010 IEEE International
  Symposium on Biomedical Imaging: From Nano to Macro},  716--719, IEEE
  (2010).

\bibitem{wang2014fast}
K.~Wang, J.~Xia, C.~Li, {\em et~al.}, ``Fast spatiotemporal image
  reconstruction based on low-rank matrix estimation for dynamic photoacoustic
  computed tomography,'' {\em Journal of biomedical optics} {\bf 19}(5),
  056007--056007  (2014).

\bibitem{arridge2016accelerated}
S.~Arridge, P.~Beard, M.~Betcke, {\em et~al.}, ``Accelerated high-resolution
  photoacoustic tomography via compressed sensing,'' {\em Physics in Medicine
  \& Biology} {\bf 61}(24), 8908  (2016).

\bibitem{compressed_razansky}
A.~Özbek, X.~L. Deán-Ben, and D.~Razansky, ``Compressed optoacoustic sensing
  of volumetric cardiac motion,'' {\em IEEE Transactions on Medical Imaging}
  {\bf 39}(10), 3250--3255  (2020).

\bibitem{lucka2018enhancing}
F.~Lucka, N.~Huynh, M.~Betcke, {\em et~al.}, ``Enhancing compressed sensing
  {4D} photoacoustic tomography by simultaneous motion estimation,'' {\em SIAM
  Journal on Imaging Sciences} {\bf 11}(4), 2224--2253  (2018).

\bibitem{wang2013accelerating}
K.~Wang, C.~Huang, Y.-J. Kao, {\em et~al.}, ``Accelerating image reconstruction
  in three-dimensional optoacoustic tomography on graphics processing units,''
  {\em Medical physics} {\bf 40}(2), 023301  (2013).

\bibitem{liang2007spatiotemporal}
Z.-P. Liang, ``Spatiotemporal imagingwith partially separable functions,'' in
  {\em 2007 4th IEEE international symposium on biomedical imaging: from nano
  to macro},  988--991, IEEE  (2007).

\bibitem{brinegar2009real}
C.~Brinegar, H.~Zhang, Y.-J.~L. Wu, {\em et~al.}, ``Real-time cardiac mri using
  prior spatial-spectral information,'' in {\em 2009 Annual International
  Conference of the IEEE Engineering in Medicine and Biology Society},
  4383--4386, IEEE  (2009).

\bibitem{cam2023dynamic}
R.~M. Cam, C.~Wang, S.~Park, {\em et~al.}, ``Dynamic image reconstruction to
  monitor tumor vascular perfusion in small animals using 3d photoacoustic
  computed-tomography imagers with rotating gantries,'' in {\em Photons Plus
  Ultrasound: Imaging and Sensing 2023},   {\bf 12379}, 78--83, SPIE  (2023).

\bibitem{cam2023low}
R.~M. Cam, C.~Wang, W.~Thompson, {\em et~al.}, ``Low-rank matrix
  estimation-based spatiotemporal image reconstruction from few tomographic
  measurements per frame for dynamic photoacoustic computed tomography,'' in
  {\em Medical Imaging 2023: Physics of Medical Imaging},   {\bf 12463},
  124630R, SPIE  (2023).

\bibitem{lozenski2022memory}
L.~Lozenski, M.~A. Anastasio, and U.~Villa, ``A memory-efficient
  self-supervised dynamic image reconstruction method using neural fields,''
  {\em IEEE Transactions on Computational Imaging} {\bf 8}, 879--892  (2022).

\bibitem{lingala2011accelerated}
S.~G. Lingala, Y.~Hu, E.~DiBella, {\em et~al.}, ``Accelerated dynamic mri
  exploiting sparsity and low-rank structure: kt slr,'' {\em IEEE transactions
  on medical imaging} {\bf 30}(5), 1042--1054  (2011).

\bibitem{gu2014weighted}
S.~Gu, L.~Zhang, W.~Zuo, {\em et~al.}, ``Weighted nuclear norm minimization
  with application to image denoising,'' in {\em Proceedings of the IEEE
  conference on computer vision and pattern recognition},  2862--2869  (2014).

\bibitem{tibshirani2010proximal}
R.~Tibshirani {\em et~al.}, ``Proximal gradient descent and acceleration,''
  {\em Lecture Notes}   (2010).

\bibitem{combettes2011proximal}
P.~L. Combettes and J.-C. Pesquet, ``Proximal splitting methods in signal
  processing,'' {\em Fixed-point algorithms for inverse problems in science and
  engineering} , 185--212  (2011).

\bibitem{beck2009fast}
A.~Beck and M.~Teboulle, ``A fast iterative shrinkage-thresholding algorithm
  for linear inverse problems,'' {\em SIAM journal on imaging sciences} {\bf
  2}(1), 183--202  (2009).

\bibitem{nesterov2013gradient}
Y.~Nesterov, ``Gradient methods for minimizing composite functions,'' {\em
  Mathematical programming} {\bf 140}(1), 125--161  (2013).

\bibitem{hudson1994accelerated}
H.~M. Hudson and R.~S. Larkin, ``Accelerated image reconstruction using ordered
  subsets of projection data,'' {\em IEEE transactions on medical imaging} {\bf
  13}(4), 601--609  (1994).

\bibitem{kim2014combining}
D.~Kim, S.~Ramani, and J.~A. Fessler, ``Combining ordered subsets and momentum
  for accelerated x-ray ct image reconstruction,'' {\em IEEE transactions on
  medical imaging} {\bf 34}(1), 167--178  (2014).

\bibitem{haase2020improved}
V.~Haase, K.~Stierstorfer, K.~Hahn, {\em et~al.}, ``An improved combination of
  ordered subsets and momentum for fast model-based iterative ct
  reconstruction,'' in {\em Medical Imaging 2020: Physics of Medical Imaging},
   {\bf 11312}, 391--397, SPIE  (2020).

\bibitem{cai2010singular}
J.-F. Cai, E.~J. Cand{\`e}s, and Z.~Shen, ``A singular value thresholding
  algorithm for matrix completion,'' {\em SIAM Journal on optimization} {\bf
  20}(4), 1956--1982  (2010).

\bibitem{halko2011finding}
N.~Halko, P.-G. Martinsson, and J.~A. Tropp, ``Finding structure with
  randomness: Probabilistic algorithms for constructing approximate matrix
  decompositions,'' {\em SIAM review} {\bf 53}(2), 217--288  (2011).

\bibitem{tritom}
H.~P. Brecht, V.~Ivanov, D.~S. Dumani, {\em et~al.}, ``{A 3D imaging system
  integrating photoacoustic and fluorescence orthogonal projections for
  anatomical, functional and molecular assessment of rodent models},'' in {\em
  Photons Plus Ultrasound: Imaging and Sensing 2018},  A.~A. Oraevsky and L.~V.
  Wang, Eds.,  {\bf 10494}, 81 -- 88, International Society for Optics and
  Photonics, SPIE  (2018).

\bibitem{thompson2023characterizing}
W.~R. Thompson, H.-P.~F. Brecht, V.~Ivanov, {\em et~al.}, ``{Characterizing a
  photoacoustic and fluorescence imaging platform for preclinical murine
  longitudinal studies},'' {\em Journal of Biomedical Optics} {\bf 28}(3),
  036001  (2023).

\bibitem{schoustra2019twente}
S.~M. Schoustra, D.~Piras, R.~Huijink, {\em et~al.}, ``{Twente Photoacoustic
  Mammoscope 2: system overview and three-dimensional vascular network images
  in healthy breasts},'' {\em Journal of Biomedical Optics} {\bf 24}(12),
  121909  (2019).

\bibitem{ito2014multi}
K.~Ito, B.~Jin, and T.~Takeuchi, ``Multi-parameter tikhonov regularization—an
  augmented approach,'' {\em Chinese Annals of Mathematics, Series B} {\bf
  35}(3), 383--398  (2014).

\bibitem{pereverzev2005adaptive}
S.~Pereverzev and E.~Schock, ``On the adaptive selection of the parameter in
  regularization of ill-posed problems,'' {\em SIAM Journal on Numerical
  Analysis} {\bf 43}(5), 2060--2076  (2005).

\bibitem{patir}
W.~Thompson, A.~Yu, D.~S. Dumani, {\em et~al.}, ``{A preclinical small animal
  imaging platform combining multi-angle photoacoustic and fluorescence
  projections into co-registered 3D maps},'' in {\em Photons Plus Ultrasound:
  Imaging and Sensing 2020},  A.~A. Oraevsky and L.~V. Wang, Eds.,  {\bf
  11240}, 112400L, International Society for Optics and Photonics, SPIE
  (2020).

\bibitem{xu2005universal}
M.~Xu and L.~V. Wang, ``Universal back-projection algorithm for photoacoustic
  computed tomography,'' {\em Physical Review E} {\bf 71}(1), 016706  (2005).

\bibitem{video}
``dynamic pact videos.''
  \url{https://uofi.box.com/s/97eqmnzctj7ctrz1rxa91gnwwg9pt8wt}.
\newblock Accessed: 2023-09-28.

\bibitem{ParaView}
J.~Ahrens, B.~Geveci, and C.~Law, {\em ParaView: An End-User Tool for Large
  Data Visualization}, 717--731.
\newblock Elsevier Butterworth-Heinemann, Burlington, MA, USA  (2005).
\newblock [doi:10.1016/B978-012387582-2/50038-1].

\bibitem{hupple2018light}
C.~W. Hupple, S.~Morscher, N.~C. Burton, {\em et~al.}, ``A
  light-fluence-independent method for the quantitative analysis of dynamic
  contrast-enhanced multispectral optoacoustic tomography (dce msot),'' {\em
  Photoacoustics} {\bf 10}, 54--64  (2018).

\end{thebibliography}
\bibliographystyle{spiejour}   




\listoffigures

\end{document}